\documentclass[aip,reprint]{revtex4-2}
\usepackage[utf8]{inputenc}
\usepackage{graphicx, color}
\usepackage{ulem}
\usepackage{xcolor}
\usepackage{hyperref}
\usepackage{amsmath}
\usepackage{amssymb}
\usepackage{mathtools}
\usepackage{float}
\usepackage{gensymb}
\usepackage{textcomp}
\usepackage{siunitx}
\usepackage{longtable}
\usepackage{booktabs}
\usepackage{natbib}
\usepackage[section]{placeins}
\usepackage{xspace}
\usepackage{url}
\usepackage[toc,page]{appendix}
\usepackage{bm}
\usepackage{float}
\usepackage[section]{placeins}
\definecolor{link}{rgb}{0.1,0.1,0.9}
\hypersetup{colorlinks=true,linkcolor=link,citecolor=link,urlcolor=link,linktocpage}

\newcommand*\pct{\protect\scalebox{0.85}{\%}\xspace}
\makeatletter
\g@addto@macro\bfseries{\boldmath}
\makeatother

\DeclareRobustCommand{\rchi}{{\mathpalette\irchi\relax}}
\newcommand{\irchi}[2]{\raisebox{\depth}{$#1\chi$}}

\newcommand{\sg}{$P$2$_1/c$\xspace}
\newcommand{\ACVO}{${A}$CoV$_{2}$O$_{7}$~(${A}$~=~Ca, Sr)\xspace}
\newcommand{\ACVOshort}{${A}$CoV$_{2}$O$_{7}$\xspace}
\newcommand{\CCVO}{CaCoV$_{2}$O$_{7}$\xspace}
\newcommand{\SCVO}{SrCoV$_{2}$O$_{7}$\xspace}
\newcommand{\CMVO}{CaMgV$_{2}$O$_{7}$\xspace}

\draft 

\begin{document}

\title{Exploring Low-Dimensional Magnetism in Cobalt Vanadates, ${A}$CoV$_{2}$O$_{7}$~(${A}$~=~Ca, Sr) : Crystal Growth and Magnetic Properties of Effective Spin-1/2 Zigzag Chains}

\author{Anzar Ali}
\email{a.ali@fkf.mpg.de}
\affiliation{Max Planck Institute for Solid State Research, Heisenbergstraße 1, D-70569 Stuttgart, Germany}

\author{Guratinder Kaur}
\affiliation{Max Planck Institute for Solid State Research, Heisenbergstraße 1, D-70569 Stuttgart, Germany}
\affiliation{School of Physics and Astronomy, University of Edinburgh, Edinburgh EH9 3JZ, United Kingdom} 

\author{Arvind Maurya}
\affiliation{Max Planck Institute for Solid State Research, Heisenbergstraße 1, D-70569 Stuttgart, Germany}
\affiliation{Department of Physics, School of Physical Sciences, Mizoram University, Aizawl 796004, India}

\author{Isha}
\affiliation{UGC-DAE Consortium for Scientific Research, University Campus, Khandwa Road, Indore-452001, India}

\author{Kathrin Küster}
\affiliation{Max Planck Institute for Solid State Research, Heisenbergstraße 1, D-70569 Stuttgart, Germany}

\author{Ulrich Starke}
\affiliation{Max Planck Institute for Solid State Research, Heisenbergstraße 1, D-70569 Stuttgart, Germany}

\author{Pascal Puphal}
\affiliation{Max Planck Institute for Solid State Research, Heisenbergstraße 1, D-70569 Stuttgart, Germany}

\author{Arvind Kumar Yogi}
\email{akyogi@csr.res.in}
\affiliation{UGC-DAE Consortium for Scientific Research, University Campus, Khandwa Road, Indore-452001, India}

\author{Masahiko Isobe}
\email{m.isobe@fkf.mpg.de}
\affiliation{Max Planck Institute for Solid State Research, Heisenbergstraße 1, D-70569 Stuttgart, Germany}

\date{\today}

\begin{abstract}
	We report the successful growth of high-quality single crystals of \ACVO, a quasi-one-dimensional zigzag chain compound containing Co$^{2+}$ ions, using the optical floating zone method. The crystal growth was stabilized under high-pressure argon-oxygen gas with slow growth rates, overcoming challenges associated with the incongruent melting behavior of this material. X-ray diffraction confirms the zigzag arrangement of Co$^{2+}$ ions, forming a quasi-one-dimensional chain structure. Magnetic susceptibility and heat capacity measurements reveal an antiferromagnetic phase transition at the N\'eel temperature ($T_{\text{N}} \sim 3.5$ K) and negative Curie-Weiss temperatures, indicative of dominant antiferromagnetic interactions. The distorted CoO$_6$ octahedral geometry and strong spin-orbit coupling suggest that Co$^{2+}$ ions likely exhibit an effective $J = 1/2 $ Kramers doublet state. The results presented here demonstrate the potential of \ACVO\ as a platform for investigating low-dimensional magnetism and quantum magnetic phenomena. These insights shed light on the role of the ${A}$-site ion in tuning the magnetic interactions, which will foster future research into the field-induced behavior in these cobalt vanadates.
	\end{abstract}

\maketitle
\section{Introduction}
\label{Intro}

Low-dimensional quantum magnets, particularly those with low spin ($S = 1/2$), offer a fertile ground for exploring exotic magnetic ground states and quantum phase transitions. These systems are characterized by quantum fluctuations that lift the degeneracy of classical antiferromagnetic (AFM) spins, driven by factors such as reduced dimensionality, low spin, geometric frustration, and external magnetic fields~\cite{Vasiliev2018, Thalmeier2024, He2005}. In such systems, the magnetic properties are governed by the interplay of symmetric and antisymmetric exchange interactions, as well as magnetocrystalline anisotropy~\cite{Shen2017, Susuki2013}. The reduction in dimensionality enhances quantum effects, making materials with one-dimensional (1D) chains or two-dimensional (2D) planes ideal platforms for studying quantum criticality and exotic quantum phases~\cite{Dagotto1999}.

Recent studies on cobalt-based quantum magnets have revealed a variety of intriguing quantum phenomena, including Bethe strings in SrCo$_2$V$_2$O$_8$ and topological quantum phase transitions in BaCo$_2$V$_2$O$_8$~\cite{Wang2018_Nature, Bera2020, Faure2018}. Compounds containing Co$^{2+}$ ($3d^7$) ions are of great interest due to their potential to host Kitaev interactions, which arise from the interplay of spin-orbit coupling (SOC) and crystal field effects (CEF)~\cite{Liu2018, Liu2020}. Honeycomb-lattice cobaltates, for instance, are promising candidates for realizing Kitaev spin liquids and other topological states. Notably, studies on triangular and honeycomb lattice cobaltates have unveiled giant magnetocaloric effects and emergent Berezinskii-Kosterlitz-Thouless (BKT) transitions, highlighting their rich magnetic behavior and potential applications~\cite{Xiang2024, Chen2024}. Theoretical models have further highlighted the richness of magnetic phases in zigzag spin chains, where competing nearest-neighbor ($J_1$) and next-nearest-neighbor ($J_2$) exchange interactions give rise to Néel and double Néel phases, separated by quantum critical points~\cite{Zhu2019, Liu2019, Sato2011}.

In particular, 1D spin-1/2 zigzag chain systems have emerged as valuable platforms to explore fractionalized excitations, spin transport, and unconventional magnetic order driven by frustration and anisotropy. Recent experimental realizations of freestanding 1D quantum spin chains have enabled the direct observation of spinon-mediated transport phenomena and quantum coherence in magnetic insulators~\cite{Hirobe2017, Li2024}. The pioneering work by Yang \textit{et al.}~\cite{Yang1966} also provides a theoretical framework for understanding anisotropic spin-spin interactions in one-dimensional systems. These developments further underscore the importance of 1D spin chain compounds for understanding quantum magnetism in low-dimensional limits and motivate the exploration of new Co-based zigzag chain materials.

Despite significant theoretical advances, experimental realizations of Co$^{2+}$-based 1D zigzag spin chains remain scarce, primarily due to the challenges of synthesizing high-quality single crystals. These challenges stem from narrow crystallization temperature ranges and specific growth conditions required for Co$^{2+}$-based materials. To date, there has been limited progress in investigating the anisotropic properties and quantum criticality in these systems due to the lack of suitable single crystals.

In this work, we present the first report on the crystal growth of the \ACVO family, including both \CCVO and \SCVO. These compounds feature 1D zigzag chains of Co$^{2+}$ ions, which adopt a pseudo-spin-$1/2$ state due to the combined effects of SOC and crystal field distortions. While a prior study by our group~\cite{Isha2024} reported field-induced quantum phase transitions and the effective $J = 1/2$ Kramers doublet state in \CCVO, the crystal growth method was not discussed. Here, we address this gap by detailing the crystal growth conditions for both \CCVO and \SCVO and providing a comprehensive study of their magnetic properties.

The significance of \ACVO compounds lies in their potential to host competing magnetic interactions and quantum fluctuations within the zigzag chain structure. Our previous work showed that \CCVO exhibits an antiferromagnetic ground state and a field-induced quantum phase transition at a critical field of $\mu_0H_{\rm c} \sim 3$~T~\cite{Isha2024}, indicative of significant spin frustration. However, comparative studies with \SCVO, particularly in the context of the role of the $A$-site ion's in tuning magnetic interactions, have been lacking. This study explores these aspects, emphasizing the structural and magnetic differences between \CCVO and \SCVO.

Through detailed magnetic susceptibility and heat capacity measurements, we confirm the antiferromagnetic transition temperatures of both compounds, with \SCVO also showing a transition around 3.5~K. Structural analysis reveals greater distortion in the Co$^{2+}$ octahedra of \CCVO, leading to stronger crystal field effects and a closer approximation to an effective spin-$J = 1/2$ configuration. These findings highlight the influence of structural parameters on magnetic interactions and quantum criticality in low-dimensional systems.

The remainder of this paper is organized as follows: Sections~\ref{Exp} and~\ref{Growth} describe the experimental details and crystal growth methods. Section~\ref{Structure} and ~\ref{XPS_S} presents the refined crystal structure and possible oxidation states of Strontium, Cobalt and Vanadium in \SCVO . Magnetic and heat capacity measurements are detailed in Sections~\ref{Mag} and~\ref{HC}. Finally, our findings are discussed in Section~\ref{Dis}, with summary in Section~\ref{Con}.

\section{Experimental Details}
\label{Exp}

Single crystals of \ACVO were grown using an optical floating zone furnace under optimized conditions of oxygen and argon gas flow ratios, pressure, and temperature, as detailed in Section~\ref{Growth}.  

The structural properties of the crystals were characterized using powder X-ray diffraction (PXRD) at room temperature. Crystals were crushed into powder and analyzed using a Rigaku MiniFlex diffractometer in Bragg–Brentano geometry with Cu K$\alpha$ radiation and a Ni filter. Rietveld refinements were carried out using the Jana2020 software suite~\cite{Jana2006}. Single-crystal X-ray diffraction (SC-XRD) data at room temperature were collected using a Rigaku XtaLAB Mini II instrument with Mo K$\alpha$ radiation. Data analysis was performed with CrysAlis (Pro), while the final refinements were carried out using Olex2 and ShelX.  

X-ray Laue diffraction patterns were obtained with a Photonic Science CCD detector and a tungsten (W) broad-spectrum X-ray source operating at 35~kV and 40~mA. The Laue patterns were indexed using the OrientExpress software~\cite{Laue}.

X-ray photoelectron spectroscopy (XPS) measurements were performed on single crystals of SrCoV$_2$O$_7$ using a Kratos Axis Ultra spectrometer equipped with a monochromated Al K$\alpha$ X-ray source ($h\nu = 1486.6$~eV). A low-energy electron flood gun was employed to mitigate surface charging. All spectra were calibrated with respect to the C~1$s$ core level of adventitious carbon at 284.8~eV.\cite{Grzegorz2024} Survey and high-resolution spectra were acquired at pass energies of 80~eV and 20~eV, respectively. Peak deconvolution was performed using a combination of Gaussian-Lorentzian line shapes with a Shirley background subtraction. For the Co 2p$_{3/2}$ and Co 2p$_{1/2}$ main peaks, the area ratio was constrained to 2:1. 

Heat capacity measurements were conducted using a Quantum Design Physical Property Measurement System (QD-PPMS). Magnetic susceptibility measurements were performed using a Magnetic Property Measurement System (MPMS XL) in both DC and RSO modes, provided by Quantum Design.

\section{Crystal Growth}
\label{Growth}
To grow single crystals of \ACVOshort, stoichiometric powders were synthesized via a solid-state reaction, following procedures previously described~\cite{Murashova1994, Murasaki2021, Isha2024}. High-purity starting materials, including CaCO$_3$ (99.999\pct, Alfa Aesar), SrCO$_3$ (99.995\pct, Sigma Aldrich), Co$_3$O$_4$ (99.9985\pct, Thermo Scientific), and V$_2$O$_5$ (99.99\pct, Thermo Scientific), were thoroughly ground and calcined at 700$^\circ$C for 12 hours. The resulting powders were pelletized and sintered at 790$^\circ$C for 12 hours. Before initiating crystal growth in an optical floating zone furnace, the phase stability of \CCVO and \SCVO under flowing Ar gas and in air was verified through powder synthesis heat treatment experiments using the solid-state reaction route. Phase purity was confirmed via powder X-ray diffraction.  

Based on these synthesis experiments, we determined that crystallizing both \CCVO and \SCVO required growth under a mixture of Ar and O$_2$ atmospheres. Initial growth attempts were conducted using:  
(i) a feed rod of the same composition in ambient conditions under flowing synthetic air or a mixture of Ar and O$_2$ gases at atmospheric pressure, and  
(ii) a feed rod under varying low and high static gas pressures, up to 8 bar, with a mixture of Ar and O$_2$ gases.  

However, these attempts were unsuccessful due to consistent challenges, including insufficient melt in the floating zone, inadequate molten liquid in the growth zone, and thinning of the molten zone, which caused detachment of the feed and seed rods. Additionally, inappropriate feed melting behavior resulted in an unstable floating zone. Adjustments to growth parameters such as lamp power, growth speed, and feed/seed rod rotation did not resolve these issues. Notably, for the \CCVO compound, using halogen lamp power exceeding 500 W at the early stages of growth further destabilized the floating zone, complicating the process.  

Single crystal growth was ultimately achieved using an optical floating zone furnace (Crystal System Corp., Model FZ-T-10000-H-III-VPR) equipped with four 300 W, 100 V halogen lamps. High-density feed and seed rods were prepared by ball-milling the sintered materials and filling them into 6 mm diameter rubber tubes. These tubes were evacuated and pressed using a Riken SMP-3 70 kN press in a stainless steel mold filled with water, producing uniform cylindrical rods (80 mm length, 6 mm diameter). The rods were heat-treated at 800$^\circ$C prior to growth.  

\begin{figure*} [!tb]
	\includegraphics[width=1.0\linewidth]{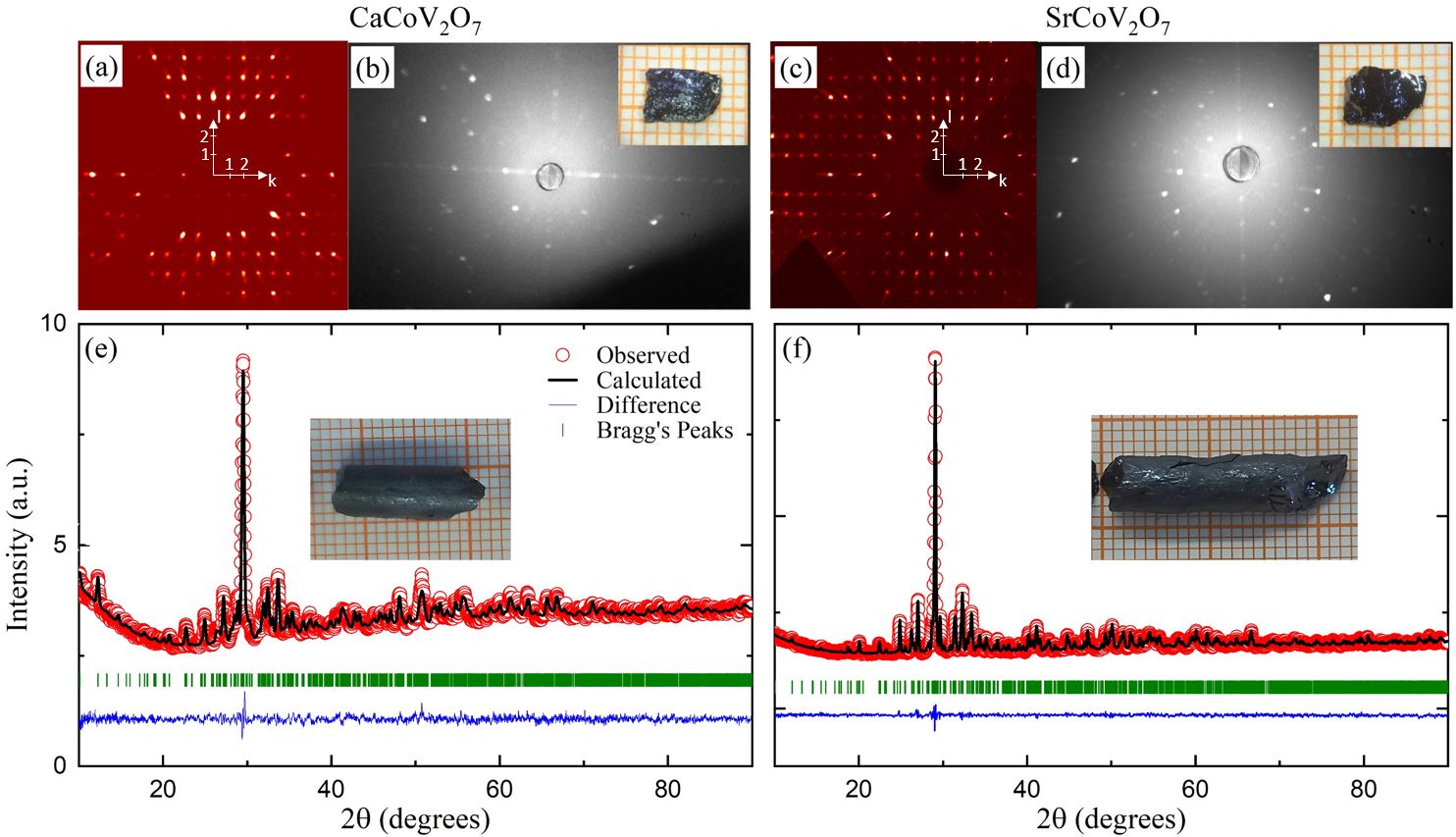}
	\caption{(a), (c) Single crystal x-ray diffraction maps of the (0$kl$) plane. (b), (d) Laue diffraction patterns obtained with the incident X-ray beam along the crystallographic $c$-axis, confirming the orientation perpendicular to the (001) plane of the crystals, shown in the insets. (e), (f) Structural refinements based on powder X-ray diffraction data at powders of crushed crystals collected at room temperature. Insets show the images of as-grown crystallized rods for \ACVO, respectively.}
	\label{Fig1}	
\end{figure*}

To optimize the growth process, we adjusted the O$_2$-to-Ar gas ratio and gas pressure within the quartz tube growth chamber. The sintered rods were pre-melted at a rate of 30 mm/h to produce O$_2$ bubble-free, dense feed rods with a final diameter of approximately 5 mm. The best single crystals were obtained under 5 bar pressure using a gas mixture of 20\pct O$_2$ and 80\pct Ar, with a growth rate of 0.3 mm/h. The rotation speeds were set to 28 rpm for the upper shaft and 24 rpm for the lower shaft. The as-grown seed rods, which displayed shiny facets (shown in the inset of Fig.~\ref{Fig1} (b, d)), yielded several crystals of varying sizes. Crystals approximately 6 mm $\times$ 4 mm $\times$ 2 mm in size were selected for Laue diffraction measurements. Using these optimized conditions, we successfully grew phase-pure single crystals of \CCVO and \SCVO, overcoming the challenges encountered in earlier trials.

\section{Results}
\label{Result}

\begin{figure*} [!tb]
	\includegraphics[width=0.90\linewidth]{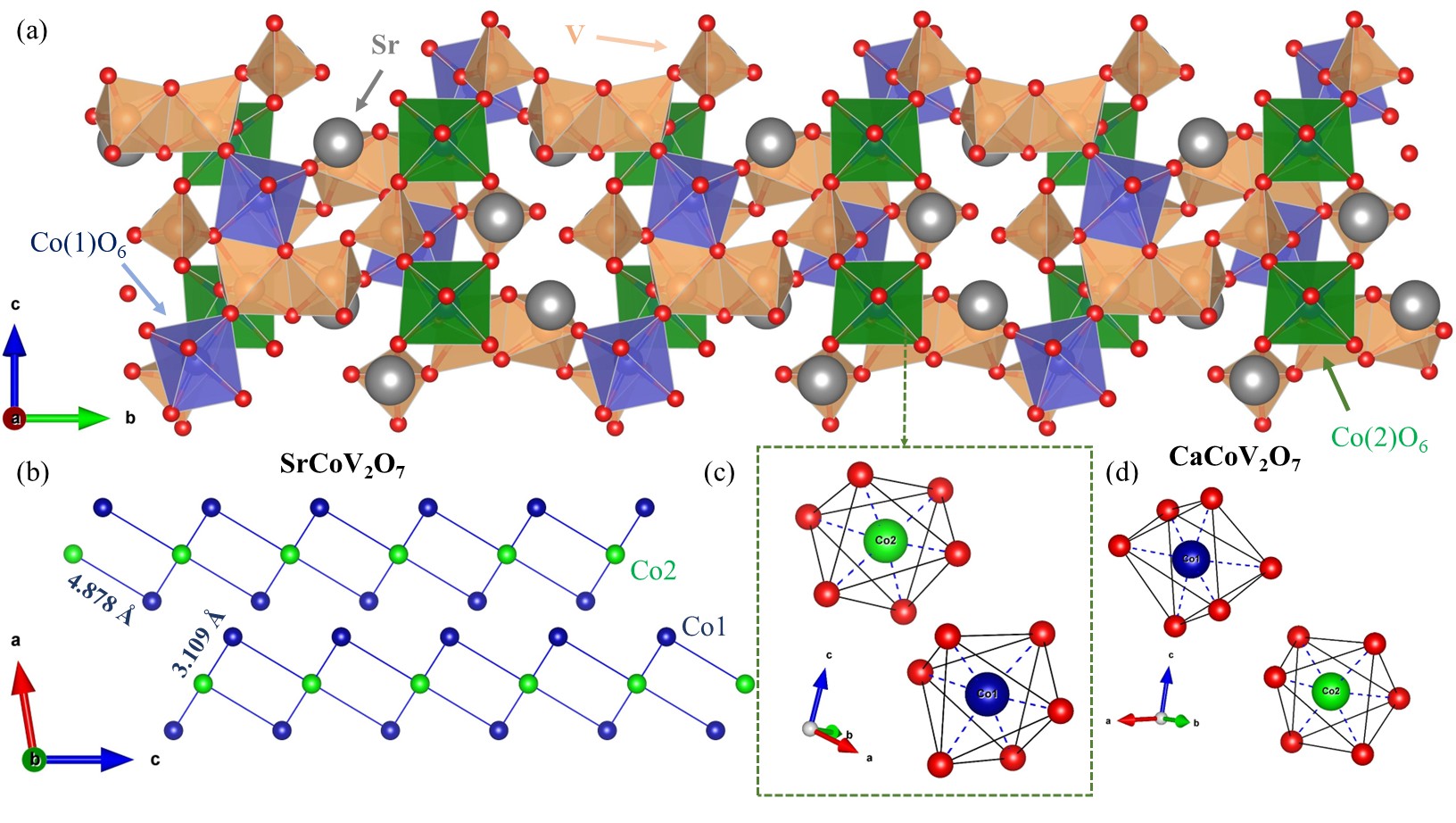}
	\caption{(a) Crystal structure of \SCVO in the $bc$-plane, showing two cobalt sites, Co(1) and Co(2), forming a zigzag chain along the crystallographic $c$-axis. (b) The spin zigzag chain of Co(1) (blue) and Co(2) (green)  along the crystallographic $c$-axis. (c), (d) Enlarged views of the distorted Co(1)O$_6$ and Co(2)O$_6$ octahedra in \ACVO, respectively.}
	\label{Fig2}	
\end{figure*}

\begin{table*}[!htb]
	
	\begin{minipage}{0.45\textwidth}	
	\caption{Refined structural parameters for \CCVO at 300~K extracted from single crystal XRD. Space group \sg, No.~14, $a$~=~6.75(12) \AA, $b$~=~14.42(12) \AA, $c$~=~11.19(87) \AA, $\beta$ = 99.72(4)$^\circ$, Goodness of fit, R factor all = 4.93\pct,  wR = 12.36\pct.}
		\label{T_1}
		\setlength\extrarowheight{4pt}
		\setlength{\tabcolsep}{3pt}
		\begin{tabular}{ccccccc}
			\hline
			Element & $x$ & $y$ & $z$ & Occ. & $U_{iso}$ & Site \\ \hline
			Ca  &  0.66505 & 1.12734 & 0.52060 & 1.000 & 0.016 & 4$e$  \\
			Ca  &  0.06571 & 0.50169 & 0.24409 & 1.000 & 0.031 & 4$e$  \\
			Co  &  0.64050 & 0.67461 & 0.57535 & 1.000 & 0.011 & 4$e$  \\
			Co  &  0.99672 & 0.25338 & 0.74708 & 1.000 & 0.011 & 4$e$  \\
			V   &  0.86767 & 0.36512 & 0.47878 & 1.000 & 0.008 & 4$e$  \\
			V   &  0.57806 & 0.51409 & 0.31218 & 1.000 & 0.011 & 4$e$  \\
			V   &  0.50948 & 0.70556 & 0.29578 & 1.000 & 0.012 & 4$e$  \\
			V   &  0.84220 & 0.87227 & 0.48560 & 1.000 & 0.010 & 4$e$  \\
			O   &  0.38690 & 0.44350 & 0.32240 & 1.000 & 0.014 & 4$e$  \\
			O   &  0.42830 & 0.60780 & 0.20520 & 1.000 & 0.016 & 4$e$  \\
			O   &  0.99360 & 0.85910 & 0.37720 & 1.000 & 0.017 & 4$e$  \\
			O   &  1.04570 & 0.34510 & 0.38570 & 1.000 & 0.017 & 4$e$  \\
			O   &  0.69130 & 0.75270 & 0.22830 & 1.000 & 0.012 & 4$e$ \\
			O   &  0.68100 & 0.28720 & 0.45040 & 1.000 & 0.014 & 4$e$  \\
			O   &  0.30790 & 0.77100 & 0.29830 & 1.000 & 0.013 & 4$e$ \\
			O   &  0.70920 & 0.47770 & 0.20910 & 1.000 & 0.018 & 4$e$  \\
			O   &  0.64510 & 0.79370 & 0.46860 & 1.000 & 0.018 & 4$e$  \\
			O   &  0.59280 & 0.62170 & 0.41090 & 1.000 & 0.013 & 4$e$  \\
			O   &  0.96700 & 0.36390 & 0.62910 & 1.000 & 0.014 & 4$e$  \\
			O   &  0.99390 & 0.86270 & 0.62270 & 1.000 & 0.020 & 4$e$  \\
			O   &  0.76760 & 0.47300 & 0.44550 & 1.000 & 0.022 & 4$e$  \\
			O   &  0.74720 & 0.97770 & 0.46560 & 1.000 & 0.035 & 4$e$  \\ \hline
		\end{tabular}
	\end{minipage}
	\hfil	
	\begin{minipage}{0.45\textwidth}
	\caption{Refined structural parameters for \SCVO at 300~K extracted from single crystal XRD. Space group \sg, No.~14, $a$~=~6.80(12) \AA, $b$~=~14.75(12) \AA, $c$~=~11.24(87) \AA, $\beta$ = 99.76(6)$^\circ$, Goodness of fit, R factor all = 6.33\pct, wR~=~12.51\pct.}
		\label{T_2}
		\setlength\extrarowheight{4pt}
		\setlength{\tabcolsep}{3pt}
		\begin{tabular}{ccccccc}
			\hline
			Element &   $x$    &   $y$   &   $z$   & Occ.  & $U_{iso}$ & Site \\ \hline
			  Sr    & -0.34370 & 0.62391 & 1.01787 & 1.000 &   0.011   & 4$e$ \\
			  Sr    & 0.91500  & 0.50490 & 0.74833 & 1.000 &   0.016   & 4$e$ \\
			  Co    & 0.36200  & 0.82443 & 0.92270 & 1.000 &   0.008   & 4$e$ \\
			  Co    & 0.00250  & 0.25322 & 0.25030 & 1.000 &   0.009   & 4$e$ \\
			   V    & 0.48570  & 0.70120 & 0.69870 & 1.000 &   0.008   & 4$e$ \\
			   V    & 0.40950  & 0.51260 & 0.67850 & 1.000 &   0.009   & 4$e$ \\
			   V    & 0.15740  & 0.63600 & 1.01930 & 1.000 &   0.007   & 4$e$ \\
			   V    & 0.12740  & 0.36230 & 0.52140 & 1.000 &   0.007   & 4$e$ \\
			   O    & 0.00000  & 0.64310 & 0.88470 & 1.000 &   0.012   & 4$e$ \\
			   O    & 0.34900  & 0.71570 & 1.03810 & 1.000 &   0.010   & 4$e$ \\
			   O    & 0.58400  & 0.43790 & 0.67060 & 1.000 &   0.015   & 4$e$ \\
			   O    & 0.57200  & 0.60150 & 0.78210 & 1.000 &   0.009   & 4$e$ \\
			   O    & 0.31650  & 0.74510 & 0.77700 & 1.000 &   0.008   & 4$e$ \\
			   O    & 0.02600  & 0.35510 & 0.37280 & 1.000 &   0.011   & 4$e$ \\
			   O    & 0.01100  & 0.64330 & 1.12910 & 1.000 &   0.011   & 4$e$ \\
			   O    & 0.31700  & 0.29010 & 0.54950 & 1.000 &   0.012   & 4$e$ \\
			   O    & -0.04400 & 0.34210 & 0.61770 & 1.000 &   0.011   & 4$e$ \\
			   O    & 0.26400  & 0.53480 & 1.03640 & 1.000 &   0.018   & 4$e$ \\
			   O    & 0.21800  & 0.47200 & 0.54610 & 1.000 &   0.012   & 4$e$ \\
			   O    & 0.39630  & 0.61790 & 0.58370 & 1.000 &   0.007   & 4$e$ \\
			   O    & 0.68800  & 0.76480 & 0.70330 & 1.000 &   0.012   & 4$e$ \\
			   O    & 0.28100  & 0.48200 & 0.78480 & 1.000 &   0.021   & 4$e$ \\ \hline
		\end{tabular}	
	\end{minipage}
\end{table*}

\subsection{Crystal Structure}
\label{Structure}

\begin{table*}
\caption{Bond lengths (\AA) and bond angles (degrees) for Co1 and Co2 atom pairings in \CCVO and \SCVO crystals.}
	\label{T_3}
	\setlength\extrarowheight{4pt}
	\setlength{\tabcolsep}{12pt}
	\begin{tabular}{cccccc}
		\hline
		\multicolumn{2}{c}{\CCVO Crystal} & \multicolumn{2}{c}{\SCVO Crystal} \\
		\hline
		Bond Lengths & Bond Angles & Bond Lengths & Bond Angles \\
		\hline
		Co1-O1 ~ 2.077(3)  & $\angle$O1-Co1-O4 ~ 87.43(3)$^\circ$ & Co1-O2 ~ 2.075(13) & $\angle$O2-Co1-O9 ~ 94.7(2)$^\circ$ \\
		Co1-O4 ~ 2.108(3)  & $\angle$O5-Co1-O6 ~ 90.20(2)$^\circ$ & Co1-O3 ~ 2.040(13) & $\angle$O5-Co1-O8 ~ 89.9(3)$^\circ$ \\
		Co1-O5 ~ 1.988(3)  & $\angle$O10-Co1-O1 ~ 100.51(5)$^\circ$ & Co1-O5 ~ 1.996(13) & $\angle$O12-Co1-O9 ~ 96.0(4)$^\circ$ \\
		Co1-O6 ~ 2.210(3)  &  & Co1-O8 ~ 2.215(14) &  \\
		Co1-O9 ~ 2.095(4)  &  & Co1-O9 ~ 2.157(14) &  \\
		Co1-O10 ~ 1.969(3) & & Co1-O12 ~ 1.980(13) &  \\
		\\ 
		Co2-O3 ~ 2.146(4)  & $\angle$O4-Co2-O3 ~ 87.09(5)$^\circ$ & Co2-O1 ~ 2.154(14) & $\angle$O6-Co2-O7 ~ 98.0(2)$^\circ$ \\
		Co2-O4 ~ 2.088(3)  & $\angle$O5-Co2-O12 ~ 91.60(3)$^\circ$ & Co2-O5 ~ 2.143(13) & $\angle$O7-Co2-O5 ~ 87.2(3)$^\circ$ \\
		Co2-O5 ~ 2.079(3)  & $\angle$O11-Co2-O3 ~ 100.36(3)$^\circ$ & Co2-O6 ~ 2.028(13) & $\angle$O13-Co2-O1 ~ 98.9(5)$^\circ$ \\
		Co2-O7 ~ 2.063(3)  &  & Co2-O7 ~ 2.125(14) &  \\
		Co2-O11 ~ 2.059(3) &  & Co2-O9 ~ 2.036(14) &  \\
		Co2-O12 ~ 2.142(4) &  & Co2-O13 ~ 2.104(15) &  \\
		\hline
	\end{tabular}
\end{table*}

The structural characterization and quality of the as-grown single crystals were assessed using single crystal X-ray diffraction, Laue backscattering, and powder X-ray diffraction. Figures~\ref{Fig1}(e, f) display the PXRD patterns measured at powders of the crushed crystals. These patterns were compared with those in the literature~\cite{Murashova1994, Zhuravlev2018, Babaryk2015}, showing excellent agreement. All reflections in the experimental pattern matched the reported \CCVO pattern, confirming the successful synthesis of phase-pure \ACVOshort.

Figures~\ref{Fig1}(a, c) show the diffraction maps of the ($0kl$) planes for \CCVO and \SCVO. Additionally, Laue diffraction measurements (Fig.~\ref{Fig1}(b, d)) were used to map the crystallographic orientation and verify the alignment of the single crystals for subsequent physical property measurements. The well-defined Laue spots indicate the high crystallinity of the as-grown samples.

Figures~\ref{Fig1}(e, f) also present the Rietveld refinement profiles derived from PXRD data for the crushed crystals. The refinements confirmed a high degree of phase purity, with no impurity peaks detected. The diffraction patterns were in excellent agreement with the expected structural model.

Both \CCVO and \SCVO crystallize in the monoclinic space group \textit{P2$_1$/c}. The lattice parameters for \CCVO were determined as \textit{a} = 6.75(12)~\AA, \textit{b} = 14.42(12)~\AA, \textit{c} = 11.19(87)~\AA, $\beta$ = 99.72(4)$^\circ$, and for \SCVO, \textit{a} = 6.80(12)~\AA, \textit{b} = 14.75(12)~\AA, \textit{c} = 11.24(87)~\AA, $\beta$ = 99.76(6)$^\circ$. These refined lattice parameters are consistent with previously reported values~\cite{Isha2024, Murashova1994, Huang2022, Babaryk2015, Zhuravlev2018}, as summarized in Tables~\ref{T_1} and \ref{T_2}.

The structural refinements revealed that both \CCVO and \SCVO contain four crystallographic sites for non-magnetic V$^{5+}$ ions and two sites each for magnetic Co$^{2+}$ and non-magnetic A$^{2+}$ ions. Oxygen ions occupy fourteen distinct crystallographic sites. All atoms are located at fully occupied 4$e$ positions.

Figures~\ref{Fig2}(a, b) depict the crystal structure of \SCVO, highlighting Co$^{2+}$ dimer chains formed by corner-sharing CoO$_6$ octahedra, distorted VO$_5$ pyramids, and nearly regular VO$_4$ tetrahedra oriented along the $c$-axis. The magnetic Co(1) and Co(2) ions are located at the centers of trigonal distorted CoO$_6$ octahedra. Vanadium ions are distributed across four distinct crystallographic sites: V(1), V(2), V(3), and V(4). Bond lengths and angles for the CoO$_6$ octahedra, derived from single crystal X-ray diffraction data, are listed in Table~\ref{T_3}. The CoO$_6$ octahedra and the VO$_5$ and VO$_4$ polyhedra exhibit deviations from ideal symmetry. Zigzag chains are interconnected within the $ab$-plane via V(1)O$_5$, V(2)O$_5$, V(3)O$_4$, and V(4)O$_4$ polyhedra. Figures~\ref{Fig2}(c, d) illustrate the distorted octahedra for \ACVOshort.

The bond angles $\angle$O-Co(1,2)-O in the CoO$_6$ octahedra show distortions, with maximum deviations from 90$^\circ$ of $\pm$10.5$^\circ$ for \CCVO and $\pm$8.9$^\circ$ for \SCVO, indicating greater distortion in \CCVO. As discussed by Isha \textit{et al.}~\cite{Isha2024}, the trigonal distortion, influenced by spin-orbit coupling, results in a pseudospin $S = 1/2$ ground state. This analysis suggests that \CCVO is closer to the effective $J_\text{eff} = 1/2$ state for Co$^{2+}$ than \SCVO.

	\subsection{X-ray Photoelectron Spectroscopy}
	\label{XPS_S}
	
	\begin{figure*}[!htb]
		\includegraphics[width=1.0\linewidth]{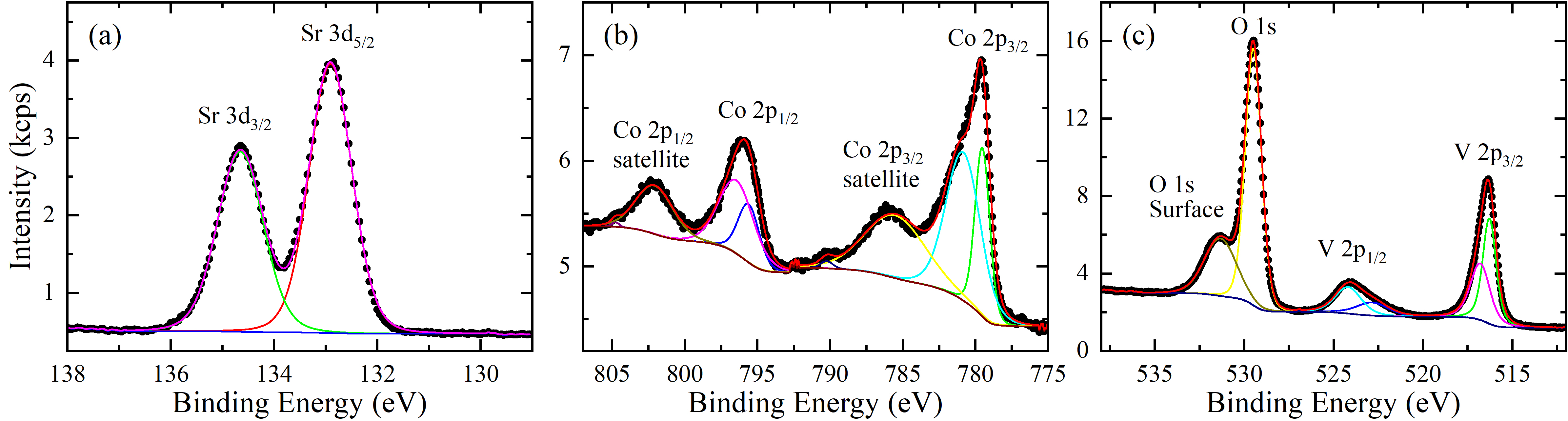}
		\caption{High-resolution X-ray photoelectron spectra of SrCoV$_2$O$_7$: (a) Sr~3$d$, (b) Co~2$p$, and (c) O~1$s$/V~2$p$ regions. Spin-orbit-split components and satellite features are indicated.}
		\label{XPS}	
	\end{figure*}
	
	Figure~\ref{XPS} displays the core-level X-ray photoelectron spectra of SrCoV$_2$O$_7$, highlighting the Sr~3$d$, Co~2$p$, and V~2$p$/O~1$s$ regions. The extracted peak parameters are summarized in Table~\ref{T_4}.
	
	The Sr~3$d$ region [Fig.~\ref{XPS}(a)] shows well-defined spin-orbit split doublets with Sr~3$d_{5/2}$ and Sr~3$d_{3/2}$ components located at 132.91~eV and 134.65~eV, respectively. The observed peak positions and narrow line widths confirm the Sr$^{2+}$ oxidation state in an oxide environment~\cite{Young1985}. No additional peaks or asymmetry is observed, suggesting the absence of mixed valence or surface carbonate contamination.
	
	In the Co~2$p$ region [Fig.~\ref{XPS}(b)], a pronounced main peak appears near 780~eV, with an evident shoulder at higher binding energy and a strong satellite structure around 785–790~eV. These features are characteristic of high-spin Co$^{2+}$ in an octahedral environment. Specifically, the Co~2$p_{3/2}$ peak appears at 779.52~eV, with a secondary component at 780.83~eV, while the Co~2$p_{1/2}$ doublet is observed at 795.59~eV and 796.44~eV. The intense satellite peaks located at 785.53~eV, 790.23~eV, 802.08~eV, and 804.77~eV are attributed to charge-transfer type screening effects. The main and satellite peaks arise from well-screened and poorly screened final states, respectively, consistent with a high-spin Co$^{2+}$ electronic configuration~\cite{Takubo2005, Isha2024, Elp1991, Groot1993}. 
	
	In contrast, a Co$^{3+}$ state typically exhibits much weaker satellite features due to stronger Co~3$d$–O~2$p$ hybridization and lower charge-transfer energy. Thus, the observed spectral weight and satellite intensity in SrCoV$_2$O$_7$ confirm the presence of Co$^{2+}$ ions. Furthermore, the full width at half maximum (FWHM) of the Co~2$p_{3/2}$ peak is found to be 1.19~eV, which is smaller than 2~eV. This suggests that the Co$^{2+}$ ions in SrCoV$_2$O$_7$ adopt an effective spin-1/2 Kramers doublet ground state, as broader FWHM values (\(\geq 2\)~eV) are typically associated with a spin-3/2 configuration~\cite{Groot1993}.
	
	The V~2$p$ spectrum [Fig.~\ref{XPS}(c)] shows two sets of spin-orbit doublets: one with V~2$p_{3/2}$ and V~2$p_{1/2}$ peaks at 516.28~eV and 522.82~eV, and another at 516.79~eV and 524.18~eV. These closely spaced features suggest the coexistence of slightly different vanadium environments, which can be attributed to the presence of both VO$_4$ tetrahedra and VO$_5$ square-pyramidal units in the crystal structure~\cite{Liardet2018, Alov2006}. Importantly, the binding energies fall within the typical range for V$^{5+}$ species, confirming the oxidation state of vanadium.
	
	Finally, the O~1$s$ spectrum reveals a dominant peak centered at 529.48~eV corresponding to lattice oxygen (O$^{2-}$), and a higher binding energy shoulder at 531.33~eV arising from surface-adsorbed species such as hydroxyl groups or carbonates. The asymmetry in the O~1$s$ envelope and the broadness of the surface-related component are typical for oxide materials exposed to air~\cite{Young1985}.
	
	Overall, the XPS data supports the formal valence state assignment of Sr$^{2+}$, Co$^{2+}$ (high-spin), and V$^{5+}$ in SrCoV$_2$O$_7$, consistent with charge balance and the proposed crystal structure. This analysis builds upon our earlier XPS study of CaCoV$_2$O$_7$~\cite{Isha2024}, which revealed comparable Co oxidation states and spectral characteristics.

	\begin{table}[h]
		\centering
		\caption{XPS fitting parameters for SrCoV$_2$O$_7$: Binding energy (B.E.) and full width at half maximum (FWHM).}
		\label{T_4}
		\setlength\extrarowheight{4pt}
		\setlength{\tabcolsep}{6pt}
		\begin{tabular}{ccc}
			\hline
			\textbf{State} & \textbf{B.E. (eV)} & \textbf{FWHM (eV)}  \\
			\hline
			Sr 3$d_{5/2}$         & 132.91  & 0.98    \\
			Sr 3$d_{3/2}$         & 134.65  & 1.05    \\
			Co 2$p_{3/2}$         & 779.52  & 1.19    \\
			Co 2$p_{1/2}$         & 795.59  & 1.73    \\
			Co 2$p_{3/2}$         & 780.83  & 2.80    \\
			Co 2$p_{1/2}$         & 796.44  & 2.84    \\
			Co 2$p_{3/2}$ sat     & 785.53  & 4.86    \\
			Co 2$p_{1/2}$ sat     & 802.08  & 3.04    \\
			Co 2$p_{3/2}$ sat     & 790.23  & 0.99    \\
			Co 2$p_{1/2}$ sat     & 804.77  & 0.60    \\
			V 2$p_{3/2}$          & 516.28  & 0.88    \\
			V 2$p_{1/2}$          & 522.82  & 2.12    \\
			V 2$p_{3/2}$          & 516.79  & 1.29    \\
			V 2$p_{1/2}$          & 524.18  & 1.71    \\
			O 1$s$                & 529.48  & 1.07    \\
			O 1$s$ (surface)      & 531.33  & 2.10    \\
			\hline
		\end{tabular}
	\end{table}

\subsection{Magnetization}
\label{Mag}

\begin{figure*} [!htb]
	\includegraphics[width=1.0\linewidth]{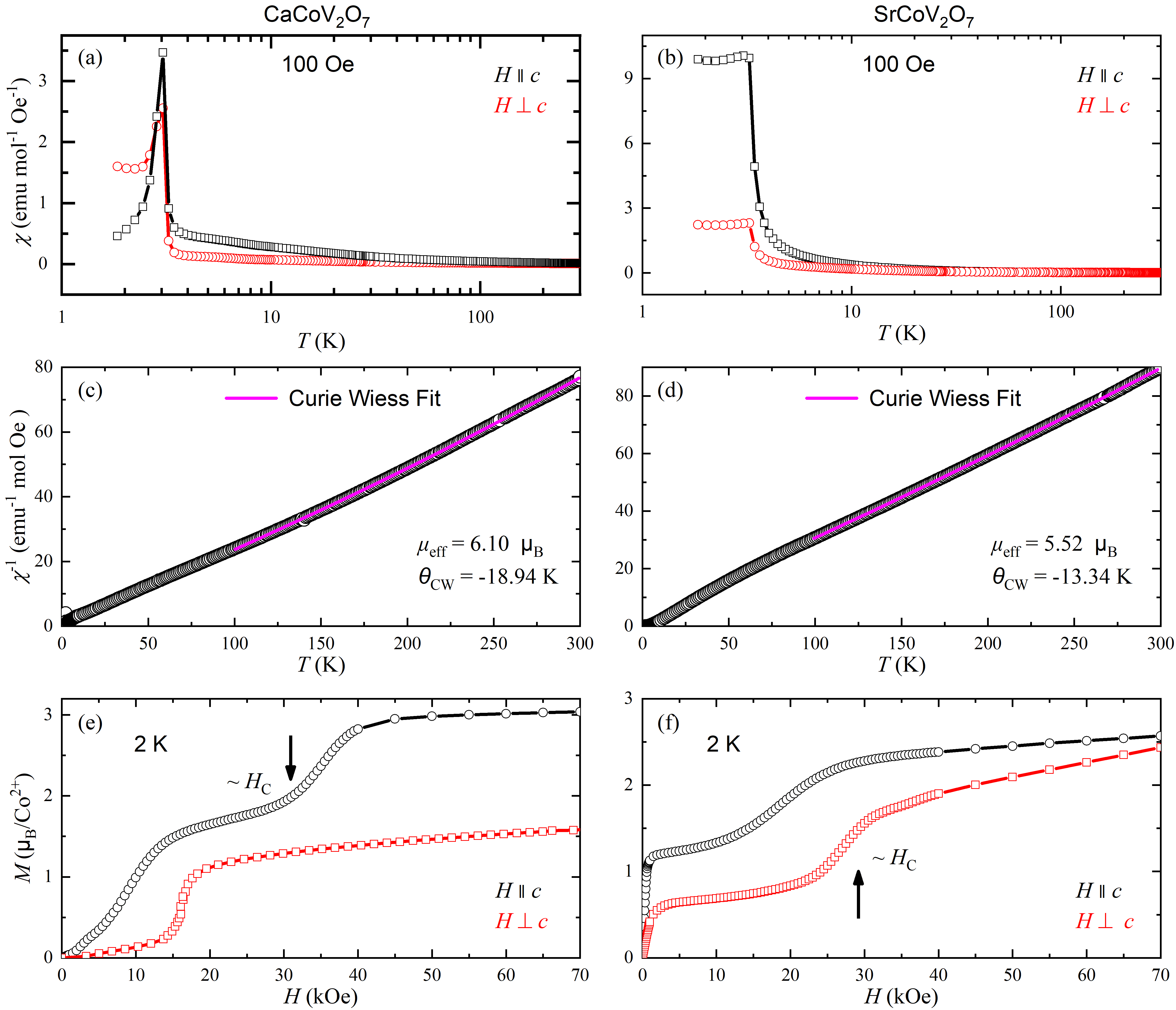}
	\caption{(a), (b) Temperature dependence of dc magnetic susceptibility at an applied field of 100 Oe along two crystallographic directions for \ACVO, respectively. (c), (d) The inverse magnetic and Curies-Weiss temperature dependence fit in the temperature range of 100-300~K for the field along the $c$-axis. (e), (f) The magnetization versus magnetic field along two crystallographic directions at a temperature of 2~K. (Black solid line for applied field along the $c$-axis, and red solid line for applied field perpendicular to the $c$-axis). 
	}
	\label{Fig3}	
\end{figure*}

Measurement of magnetic susceptibility, $\rchi(T)$, was conducted to investigate the magnetic properties of both \ACVO\ compounds. The temperature-dependent susceptibility data measured under an applied magnetic field of $H = 100 \, \text{Oe}$ are presented in Figs.~\ref{Fig3}(a) and \ref{Fig3}(b), with $H$ aligned parallel (black solid line) and perpendicular (red solid line) to the $c$-axis, covering the range from 300~K to 2~K. The susceptibility increases as the temperature decreases, with a pronounced long-range ordering transition observed at $T_\text{N} \sim~3.5 \, ~\text{K}$ for both \CCVO\ and \SCVO, as shown in Figs.~\ref{Fig3}(a) and \ref{Fig3}(b). This sharp feature signals the onset of AFM ordering. Notably, the substantial difference in susceptibility between $H \parallel c$ and $H \perp c$ highlights significant magnetic anisotropy in both compounds.

At higher temperatures ($T > 100 \, \text{K}$), $\rchi(T)$ adheres to the Curie-Weiss law, as evidenced by the inverse susceptibility data in Figs.~\ref{Fig3}(c) and \ref{Fig3}(d). These curves were fitted with the Curie-Weiss expression in the temperature range of 100~K to 300~K for $H \parallel c$: 
\[
\rchi(T) \equiv \rchi_0 + \frac{C}{T - \theta_{\text{CW}}},
\]
where $\rchi_{0}$ is the temperature-independent susceptibility, $C$ is the Curie constant, and $\theta_{\text{CW}}$ is the Curie-Weiss  temperature.
The effective magnetic moment was calculated using $\mu_\textrm{eff}$ = $\sqrt{3k_{B}C/N_{A}}$~\cite{Mugiraneza2022}, where $N_{A}$ is the Avogadro constant and $k_{B}$ is the Boltzmann constant.

The fitting results (100~–~300~K) yielded $\rchi_0 = -2.28(5) \times 10^{-3} \, ~\text{emu} \, \text{Oe}^{-1} \, \text{mol}^{-1}$, $C = 4.66(2) \, ~\text{emu} \, \text{Oe}^{-1} \, \text{mol}^{-1} \, \text{K}$, and $\theta_{\text{CW}} = -18.94(5) \, ~\text{K}$ for \CCVO, and $\rchi_0 = -1.05(2) \times 10^{-3} \, ~\text{emu} \, \text{Oe}^{-1} \, \text{mol}^{-1}$, $C = 3.81(1) \, ~\text{emu} \, \text{Oe}^{-1} \, \text{mol}^{-1} \, \text{K}$, and $\theta_{\text{CW}} = -13.34(3) \, ~\text{K}$ for \SCVO. The effective magnetic moments were $\mu_{\text{eff}} = 6.10 \, \mu_B$ for \CCVO\ and $\mu_{\text{eff}} = 5.52 \, \mu_B$ for \SCVO. The negative $\theta_{\text{CW}}$ values suggest dominant antiferromagnetic interactions.  

The magnetic properties are primarily dictated by the Co$^{2+}$ ions, as A$^{2+}$ and V$^{5+}$ are non-magnetic. For Co$^{2+}$ ($d^7$), the expected effective moments are 3.88~$\mu_B$ and 1.73~$\mu_B$ for high- and low-spin states, respectively. The observed effective moments, however, deviate from these values due to the interplay of CEF and SOC. This interaction results in a pseudo-spin-$1/2$ ground state for Co$^{2+}$, consistent with previous studies on Co$^{2+}$ octahedral systems such as Na$_2$Co$_2$TeO$_6$ \cite{Lin2021, Kim2022} and BaCo$_2$(AsO$_4$)$_2$ \cite{Zhang2023}.

The effective moment for \CCVO\ matches prior reports by Isha et al. \cite{Isha2024}, where inelastic neutron scattering revealed a pseudo-spin-$1/2$ configuration with an anisotropic Landé $g$-factor ($g_Z = 5.35$) obtained from ESR spectra. For a spin-$1/2$ model, the calculated $g$-values are $g = 7.05$ for \CCVO\ and $g = 6.37$ for \SCVO, closely aligning with ESR and Bonner-Fisher analyses for \CCVO \cite{Isha2024}.

In both compounds, the frustration factor, $|\theta_{\text{CW}}| / T_{\text{N}} \sim 5.4~(3.8)$ for \CCVO (\SCVO), indicates moderate frustration due to competing nearest-neighbor and next-nearest-neighbor interactions within the zigzag spin chain lattice. The magnetization, $M$, as a function of magnetic field, $H$, for both crystallographic orientations (0 to 70~kOe) is presented in Figs.~\ref{Fig3}(e) and \ref{Fig3}(f). At 2~K, both CaCoV$_2$O$_7$ and SrCoV$_2$O$_7$ exhibit a subtle low-field anomaly, followed by a more pronounced field-induced transition near 30~kOe, and a gradual approach toward saturation at higher fields. 

Notably, the isothermal magnetization does not fully saturate up to 70~kOe for either compound in any orientation. In both cases, the magnetization is larger along the $c$-axis---which aligns with the zigzag spin chain direction---indicating strong magnetic anisotropy and identifying the $c$-axis as the magnetic easy axis. The $M$--$H$ curves reveal a slope change below 20~kOe for \CCVO and below 2~kOe for \SCVO, followed by a more distinct anomaly near $H_{\rm c} \sim 30$~kOe. This field-induced transition in CaCoV$_2$O$_7$ has previously been identified as a quantum phase transition based on thermodynamic and neutron scattering data~\cite{Isha2024}. A comparable feature is also present in SrCoV$_2$O$_7$, though its nature remains to be established.
For the case of CaCoV$_2$O$_7$, this quantum phase transition is observed at $H_{\rm c} \sim 30$~kOe only for $H \parallel c$, whereas in SrCoV$_2$O$_7$, we observe a field-induced transition feature for both $H \parallel c$ and $H \perp c$. These contrasting behaviors suggest distinct natures of the field-induced phase transitions in the two compounds. The replacement of Sr by the smaller Ca ion on the $A$-site likely modifies the local crystal field environment and induces different distortions of the CoO$_6$ octahedra. These structural distortions can lead to changes in orbital overlap and affect the strength and anisotropy of exchange interactions, thereby influencing the spin dynamics and magnetic ground states. In particular, such differences may give rise to anisotropic responses of the spin system under external magnetic fields, as reflected in the orientation-dependent behavior of the field-induced transitions.

The incomplete saturation likely arises from strong exchange interactions and a field-driven evolution of the spin state. The origin of the low-field anomalies in both compounds is currently unclear, but may signal additional field-induced phases or spin reorientations. To clarify these features, future neutron diffraction studies under applied magnetic fields with fine field increments will be essential. Complementary low-temperature techniques, such as $\mu$SR, would also be valuable in uncovering the underlying magnetic structure.

Temperature-dependent $dc$-susceptibility measurements (not shown), performed at $\mu_0 H = 0.01$, 0.1, and 1~T, further emphasize the strong magnetic anisotropy and field sensitivity. The long-range ordering anomaly becomes progressively suppressed with increasing field, highlighting the intricate interplay of anisotropy, magnetic frustration, and external field effects in the \ACVO\ compounds.

\subsection{Heat Capacity}
\label{HC}

\begin{figure*} [!htb]
	\includegraphics[width=\linewidth]{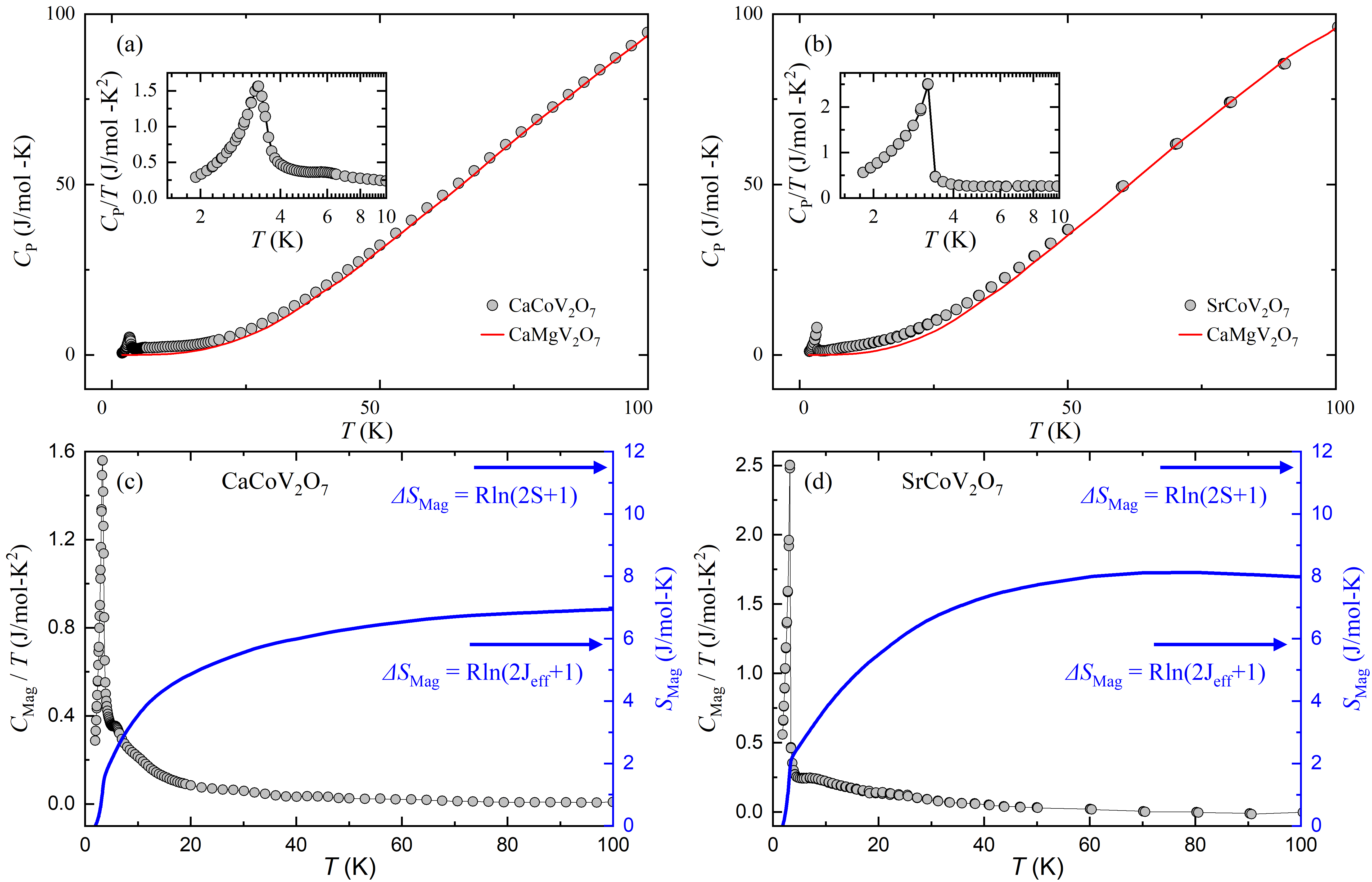}
	\caption{(a), (b) Temperature dependence of specific heat capacity for \ACVO (open symbols) and \CMVO (solid red line) measured in zero fields over the temperature range 2-100 K.  Insets show an enlarged view of the $\lambda$-like anomaly in the normalized heat capacity $C_{\text{P}}/T$ in the temperature range 2-10 K. (c), (d) The normalized magnetic entropy $C_{\text{Mag}}/T$ on the left $y$-axis and the change in magnetic entropy $\Delta$S$_{\text{Mag}}$ on the right $y$-axis. The magnetic heat capacity for both materials has been calculated by subtracting the rescaled heat capacity of non-magnetic \CMVO from the total heat capacity of \ACVOshort.
	}
	\label{Fig4}	
\end{figure*}

To further investigate the thermodynamics of the AFM transition and extract the magnetic entropy associated with the Co$^{2+}$ spins in \ACVO, heat capacity measurements ($C_{\rm p}(T)$) were performed on single crystals of \CCVO\ and \SCVO. These measurements aimed to confirm the magnetic ordering suggested by susceptibility data and provide insights into the spin degrees of freedom.

The low-temperature specific heat, $C_{\rm p}(T)$, measured via the thermal-relaxation technique at zero field, is presented in Fig.~\ref{Fig4} for both \CCVO\ and \SCVO. At high temperatures, $C_{\rm p}(T)$ is dominated by phonon excitations \cite{Kittel2004}. Between 2~K and 100~K, $C_{\rm p}(T)$ data reveal a $\lambda$-type anomaly at $T_{\rm N} \sim~ 3.5 \, \text{~K}$, indicating long-range magnetic ordering, consistent with the magnetization results. Below $T_{\rm N}$, the heat capacity gradually decreases toward zero, dominated by the magnetic contribution below the AFM transition. This behavior aligns with spin-$1/2$ zigzag chain antiferromagnetism in these materials~\cite{Hase1993, M2013}. The insets of Figs.~\ref{Fig4}(a) and \ref{Fig4}(b) provide an enlarged view of $C_{\rm P}(T)/T$ at lower temperatures, highlighting these critical features.

To extract the magnetic contribution to the heat capacity, the nonmagnetic analog \CMVO\ was employed as a reference, using the same heat capacity data as in Isha et al. \cite{Isha2024}. Since \CMVO\ has a similar monoclinic crystal structure \cite{Murashova1994} but lacks magnetic ions, it serves as a suitable proxy for estimating the lattice heat capacity of \CCVO\ and \SCVO. To account for the differences in formula masses between \CMVO\ and the \ACVO\ compounds, the lattice contribution from \CMVO\ was adjusted by rescaling the temperature, $T$, according to the relation:

\begin{equation}
	T^* = \frac{T}{\left( \frac{{\rm M}_{\text{ACoV}_2\text{O}_7}}{{\rm M}_{\text{CaMgV}_2\text{O}_7}} \right)^{1/2}},
	\label{eq1}
\end{equation}

\noindent as derived from the Debye model, where the Debye temperature $\theta_{\text{D}}$ scales with the inverse square root of the formula mass, $\theta_{\text{D}} \sim M^{-1/2}$ or $T/\theta_{\text{D}} \sim M^{1/2}$ \cite{Kittel2004}.

The rescaled lattice heat capacity for \CMVO, shown as solid red lines in Figs.~\ref{Fig4}(a) and \ref{Fig4}(b), was subtracted from the measured $C_{\rm P}(T)$ data of \CCVO\ and \SCVO\ to obtain the magnetic heat capacity, $C_{\rm mag}(T)$. This subtraction reveals a sharp peak in $C_{\rm mag}(T)/T$ near $T_{\rm N}$, as depicted in Figs.~\ref{Fig4}(c) and \ref{Fig4}(d), confirming the antiferromagnetic transitions. Interestingly, $C_{\rm mag}(T)$ does not immediately drop to zero above $T_{\rm N}$; instead, it declines gradually, becoming negligible above 50~K. This nonzero $C_{\rm mag}(T)$ above $T_{\rm N}$ reflects the presence of short-range AFM correlations extending well beyond the Néel temperature~\cite{Hase1993, Okamoto2007, Bera2016}.

Furthermore, the magnetic entropy was calculated by integrating $C_{\rm mag}(T)/T$ up to 100~K \cite{Yogi2019}:

\begin{equation}
	S_{\rm mag}(T) = \int_{T_0}^{T} \frac{C_{\rm mag}(T')}{T'} \, dT',
	\label{eq2}
\end{equation}

\noindent where $T_0$ is the base temperature. The entropy increases significantly from 2~K to approximately 50~K, reaching about 6.98~(7.95)~J/mol/K per Co\(^{2+}\) ion for \CCVO\ (\SCVO). These values deviate significantly from those expected for a spin $S = 3/2$ ($\Delta$S$_{\text{Mag}}$ = \(R\ln4 = 11.52 \, \text{J/mol/K}\)) state, instead aligning more closely with the reduced entropy expected for a pseudo-spin-$1/2$ Co$^{2+}$ ($3d^7$) ion, which is indicative of a Kramers-active state. The magnetic entropy up to the AFM phase transition accounts for only about 30\pct of \(R\ln2\), indicating strong spin correlations above \(T_{\rm N}\), characteristic of frustrated magnetism. Notably, the saturated magnetic entropy for \CCVO\ is closer to the theoretical limit ($\Delta$S$_{\text{Mag}}$ = \(R\ln2 = 5.76 \, ~\text{J/mol/K}\)) for a pseudo-spin-$1/2$ system compared to \SCVO, suggesting stronger CEF and SOC interactions in \CCVO. The pseudo-spin-$1/2$ ground state arises from the intricate interplay between CEF and SOC, which splits the $d$-orbitals and quenches orbital contributions to the entropy. Similar behavior is observed in related systems such as Na$_2$Co$_2$TeO$_6$ \cite{Lin2021, Kim2022} and BaCo$_2$(AsO$_4$)$_2$ \cite{Zhang2023}.

\section{Discussion}
\label{Dis}

We discuss the results of single-crystal X-ray diffraction analysis, XPS, magnetic susceptibility, and heat capacity measurements for \CCVO{} and \SCVO. Previous studies \cite{Isha2024} reported that \CCVO{} behaves as a pseudo-spin-1/2 zigzag chain antiferromagnet, undergoing a field-induced quantum phase transition at a moderate field of 3 T. In this work, we extend the exploration of magnetic properties to \SCVO{}, emphasizing the role of trigonal distortion, which enhances the CEF and SOC, resulting in a $J = 1/2$ Co$^{2+}$ Kramers doublet.

Among 3$d$ transition metal ions, cobalt (Co) is particularly intriguing due to its multiple spin states. The spin state of cobalt is determined by a complex interplay of factors, including its oxidation state, the surrounding CEF, coordination environment, local symmetry, and the influence of neighboring ions. This reflects a delicate balance between the CEF and Hund's rule exchange interaction. Additionally, cobalt's spin and orbital moments are coupled via SOC ($\lambda$), where $\lambda > 0$. In both \CCVO{} and \SCVO, Co$^{2+}$ ions occupy trigonally distorted CoO$_6$ octahedral sites, which favor a pseudo spin-1/2 state.

Phase-pure single crystals of both \CCVO{} and \SCVO{} were successfully grown using the optical floating zone technique. A higher partial pressure of Ar and O$_2$ gas mixtures was instrumental in stabilizing the molten zone, preventing air bubble formation caused by the decomposition of CoO and V$_2$O$_5$. Controlled mixtures of Ar and O$_2$ gases, with optimized partial pressure, enabled steady crystal growth, resulting in high-quality crystals. Initial attempts under ambient conditions or in an Ar-only atmosphere were unsuccessful.

Both \CCVO{} and \SCVO{} crystallize in the monoclinic space group \sg, as confirmed by single-crystal X-ray diffraction. Their crystal structures consist of CoO$_6$ octahedra forming zigzag spin chains, which share corners with VO$_5$ pyramids and VO$_4$ tetrahedra along the crystallographic \textit{c}-axis. The CoO$_6$ octahedra are distorted, with bond angle deviations $\angle$O-Co(1,2)-O from 90$^\circ$ reaching $\pm 10.5^\circ$ in \CCVO{} and $\pm 8.9^\circ$ in \SCVO{}, indicating greater structural distortion in \CCVO{}. 

Magnetic susceptibility measurements reveal AFM ordering near $T_{\rm N} \sim~ 3.5$ K, with notable magnetic anisotropy. Effective magnetic moments from Curie-Weiss fits in the high-temperature region are 6.10 $\mu_B$ for \CCVO{} and 5.52 $\mu_B$ for \SCVO{}, consistent with a pseudo $J = 1/2$ Co$^{2+}$ configuration arising from strong CEF and SOC \cite{Isha2024}. For \CCVO{}, ESR spectroscopy reveals highly anisotropic Landé \textit{g}-factors, with $g_x = 1.85$, $g_y = 2.17$, and $g_z = 5.35$, consistent with a prior Bonner-Fisher fit~\cite{Isha2024}. While direct ESR measurements for \SCVO are not currently available, the effective g-value estimated from Curie-Weiss analysis suggests a similar, albeit slightly weaker, spin-orbit coupled J= 1/2  ground state.

Heat capacity measurements show a lambda-like anomaly at $T_{\rm N} \sim~ 3.5$ K, confirming long-range AFM ordering. Magnetic entropy, calculated between 2 K and 100 K, supports an effective $J = 1/2$ state rather than an $S = 3/2$ state. Notably, the magnetic entropy for \CCVO{} is closer to the $J = 1/2$ limit, likely due to its greater distortion of CoO$_6$ octahedra.

Our findings underscore the need for further investigation into the distinct magnetic properties of \CCVO{} and \SCVO{}. The field-dependent neutron diffraction experiments below ~2~T with fine field steps would be highly beneficial in probing the nature of intermediate region in low mangetic field. Single-crystal inelastic neutron scattering will be essential for probing the effective spin state of Co$^{2+}$ and quantifying exchange interactions through comparison with theoretical models. Complementary ESR measurements on \SCVO{} single crystals will help determine the anisotropic Landé \textit{g}-factors and assess spin-orbit contributions. Beyond \ACVO{}, the broader zigzag spin chain family presents a compelling platform for exploring the interplay between trigonal distortion, effective spin states, and field-induced quantum phase transitions.

\section{Summary}
\label{Con}

In this study, we successfully grew large single crystals of \CCVO{} and \SCVO{} using the optical floating zone method. This technique allowed us to establish optimized growth conditions for producing high-quality single crystals of both compounds. Structural characterization via single-crystal X-ray diffraction, Laue backscattering, powder XRD, and XPS confirmed that \CCVO{} and \SCVO{} crystallize in an isostructural one-dimensional (1D) zigzag spin-chain arrangement, composed of magnetic Co$^{2+}$ ions. The trigonally distorted CoO$_6$ octahedra are likely to generate a crystal electric field for the $S = 3/2$ and $L = 1$ manifold, which, under spin-orbit interaction, could result in a pseudo spin-1/2 ground state.

Our results suggest that quasi-one-dimensional spin-chain structures play a significant role in influencing the observed magnetic behaviors. Magnetic susceptibility measurements revealed a downturn in susceptibility below $T_{\rm N}$, accompanied by negative Curie-Weiss temperatures, indicative of AFM ordering. Heat capacity measurements further supported these findings, showing clear anomalies near $T_{\rm N}$ that confirm long-range magnetic ordering in both compounds.

The intriguing magnetic properties of \CCVO and \SCVO, driven by their effective spin-1/2 one-dimensional zigzag spin chains, suggest that these materials may serve as platforms to explore field-induced phenomena, such as quantum phase transitions. The high-quality single crystals obtained in this study provide a platform for future investigations into the magnetic phase diagrams and quantum phenomena in these materials. The present study of the structure and magnetism in \ACVO{} compounds has revealed intriguing magnetic phenomena that provide an opportunity to explore quantum criticality and field-induced behaviors in quasi-one-dimensional spin systems.

\section*{Acknowledgments}
We would like to thank Eva Brücher for her valuable assistance with the magnetization and heat capacity measurements using the Quantum Design MPMS-XL and PPMS systems at Max Planck Institute for Solid State Research. The DST-SERB supported the work at the UGC-DAE Consortium for Scientific Research Indore through CRG/2022/005666.
\section*{Data Availability Statement}
The data that support the findings of this study are available from the corresponding author upon reasonable request.


\begin{thebibliography}{44}%
	\makeatletter
	\providecommand \@ifxundefined [1]{%
		\@ifx{#1\undefined}
	}%
	\providecommand \@ifnum [1]{%
		\ifnum #1\expandafter \@firstoftwo
		\else \expandafter \@secondoftwo
		\fi
	}%
	\providecommand \@ifx [1]{%
		\ifx #1\expandafter \@firstoftwo
		\else \expandafter \@secondoftwo
		\fi
	}%
	\providecommand \natexlab [1]{#1}%
	\providecommand \enquote  [1]{``#1''}%
	\providecommand \bibnamefont  [1]{#1}%
	\providecommand \bibfnamefont [1]{#1}%
	\providecommand \citenamefont [1]{#1}%
	\providecommand \href@noop [0]{\@secondoftwo}%
	\providecommand \href [0]{\begingroup \@sanitize@url \@href}%
	\providecommand \@href[1]{\@@startlink{#1}\@@href}%
	\providecommand \@@href[1]{\endgroup#1\@@endlink}%
	\providecommand \@sanitize@url [0]{\catcode `\\12\catcode `\$12\catcode
		`\&12\catcode `\#12\catcode `\^12\catcode `\_12\catcode `\%12\relax}%
	\providecommand \@@startlink[1]{}%
	\providecommand \@@endlink[0]{}%
	\providecommand \url  [0]{\begingroup\@sanitize@url \@url }%
	\providecommand \@url [1]{\endgroup\@href {#1}{\urlprefix }}%
	\providecommand \urlprefix  [0]{URL }%
	\providecommand \Eprint [0]{\href }%
	\providecommand \doibase [0]{https://doi.org/}%
	\providecommand \selectlanguage [0]{\@gobble}%
	\providecommand \bibinfo  [0]{\@secondoftwo}%
	\providecommand \bibfield  [0]{\@secondoftwo}%
	\providecommand \translation [1]{[#1]}%
	\providecommand \BibitemOpen [0]{}%
	\providecommand \bibitemStop [0]{}%
	\providecommand \bibitemNoStop [0]{.\EOS\space}%
	\providecommand \EOS [0]{\spacefactor3000\relax}%
	\providecommand \BibitemShut  [1]{\csname bibitem#1\endcsname}%
	\let\auto@bib@innerbib\@empty
	\bibitem [{\citenamefont {Vasiliev}\ \emph {et~al.}(2018)\citenamefont
		{Vasiliev}, \citenamefont {Volkova}, \citenamefont {Zvereva},\ and\
		\citenamefont {Markina}}]{Vasiliev2018}%
	\BibitemOpen
	\bibfield  {author} {\bibinfo {author} {\bibfnamefont {A.}~\bibnamefont
			{Vasiliev}}, \bibinfo {author} {\bibfnamefont {O.}~\bibnamefont {Volkova}},
		\bibinfo {author} {\bibfnamefont {E.}~\bibnamefont {Zvereva}},\ and\ \bibinfo
		{author} {\bibfnamefont {M.}~\bibnamefont {Markina}},\ }\bibfield  {title}
	{\enquote {\bibinfo {title} {{Milestones of low-D quantum magnetism}},}\
	}\href {https://doi.org/10.1038/s41535-018-0090-7} {\bibfield  {journal}
		{\bibinfo  {journal} {npj Quantum Materials}\ }\textbf {\bibinfo {volume}
			{3}},\ \bibinfo {pages} {18} (\bibinfo {year} {2018})}\BibitemShut {NoStop}%
	\bibitem [{\citenamefont {Thalmeier}\ and\ \citenamefont
		{Akbari}(2024)}]{Thalmeier2024}%
	\BibitemOpen
	\bibfield  {author} {\bibinfo {author} {\bibfnamefont {P.}~\bibnamefont
			{Thalmeier}}\ and\ \bibinfo {author} {\bibfnamefont {A.}~\bibnamefont
			{Akbari}},\ }\bibfield  {title} {\enquote {\bibinfo {title} {{Induced quantum
					magnetism in crystalline electric field singlet ground state models:
					Thermodynamics and excitations}},}\ }\href
	{https://doi.org/10.1103/PhysRevB.109.115110} {\bibfield  {journal} {\bibinfo
			{journal} {Phys. Rev. B}\ }\textbf {\bibinfo {volume} {109}},\ \bibinfo
		{pages} {115110} (\bibinfo {year} {2024})}\BibitemShut {NoStop}%
	\bibitem [{\citenamefont {He}\ \emph {et~al.}(2005)\citenamefont {He},
		\citenamefont {Taniyama}, \citenamefont {Ky\^omen},\ and\ \citenamefont
		{Itoh}}]{He2005}%
	\BibitemOpen
	\bibfield  {author} {\bibinfo {author} {\bibfnamefont {Z.}~\bibnamefont
			{He}}, \bibinfo {author} {\bibfnamefont {T.}~\bibnamefont {Taniyama}},
		\bibinfo {author} {\bibfnamefont {T.}~\bibnamefont {Ky\^omen}},\ and\
		\bibinfo {author} {\bibfnamefont {M.}~\bibnamefont {Itoh}},\ }\bibfield
	{title} {\enquote {\bibinfo {title} {{Field-induced order-disorder transition
					in the quasi-one-dimensional anisotropic antiferromagnet
					$\mathrm{Ba}{\mathrm{Co}}_{2}{\mathrm{V}}_{2}{\mathrm{O}}_{8}$}},}\ }\href
	{https://doi.org/10.1103/PhysRevB.72.172403} {\bibfield  {journal} {\bibinfo
			{journal} {Phys. Rev. B}\ }\textbf {\bibinfo {volume} {72}},\ \bibinfo
		{pages} {172403} (\bibinfo {year} {2005})}\BibitemShut {NoStop}%
	\bibitem [{\citenamefont {Shen}\ \emph {et~al.}(2017)\citenamefont {Shen},
		\citenamefont {Jellyman}, \citenamefont {Forgan}, \citenamefont {Blackburn},
		\citenamefont {Laver}, \citenamefont {Can\'evet}, \citenamefont {Schefer},
		\citenamefont {He},\ and\ \citenamefont {Itoh}}]{Shen2017}%
	\BibitemOpen
	\bibfield  {author} {\bibinfo {author} {\bibfnamefont {L.}~\bibnamefont
			{Shen}}, \bibinfo {author} {\bibfnamefont {E.}~\bibnamefont {Jellyman}},
		\bibinfo {author} {\bibfnamefont {E.~M.}\ \bibnamefont {Forgan}}, \bibinfo
		{author} {\bibfnamefont {E.}~\bibnamefont {Blackburn}}, \bibinfo {author}
		{\bibfnamefont {M.}~\bibnamefont {Laver}}, \bibinfo {author} {\bibfnamefont
			{E.}~\bibnamefont {Can\'evet}}, \bibinfo {author} {\bibfnamefont
			{J.}~\bibnamefont {Schefer}}, \bibinfo {author} {\bibfnamefont
			{Z.}~\bibnamefont {He}},\ and\ \bibinfo {author} {\bibfnamefont
			{M.}~\bibnamefont {Itoh}},\ }\bibfield  {title} {\enquote {\bibinfo {title}
			{{Unconventional magnetic phase separation in
					$\ensuremath{\gamma}{\text{-CoV}}_{2}{\mathrm{O}}_{6}$}},}\ }\href
	{https://doi.org/10.1103/PhysRevB.96.054420} {\bibfield  {journal} {\bibinfo
			{journal} {Phys. Rev. B}\ }\textbf {\bibinfo {volume} {96}},\ \bibinfo
		{pages} {054420} (\bibinfo {year} {2017})}\BibitemShut {NoStop}%
	\bibitem [{\citenamefont {Susuki}\ \emph {et~al.}(2013)\citenamefont {Susuki},
		\citenamefont {Kurita}, \citenamefont {Tanaka}, \citenamefont {Nojiri},
		\citenamefont {Matsuo}, \citenamefont {Kindo},\ and\ \citenamefont
		{Tanaka}}]{Susuki2013}%
	\BibitemOpen
	\bibfield  {author} {\bibinfo {author} {\bibfnamefont {T.}~\bibnamefont
			{Susuki}}, \bibinfo {author} {\bibfnamefont {N.}~\bibnamefont {Kurita}},
		\bibinfo {author} {\bibfnamefont {T.}~\bibnamefont {Tanaka}}, \bibinfo
		{author} {\bibfnamefont {H.}~\bibnamefont {Nojiri}}, \bibinfo {author}
		{\bibfnamefont {A.}~\bibnamefont {Matsuo}}, \bibinfo {author} {\bibfnamefont
			{K.}~\bibnamefont {Kindo}},\ and\ \bibinfo {author} {\bibfnamefont
			{H.}~\bibnamefont {Tanaka}},\ }\bibfield  {title} {\enquote {\bibinfo {title}
			{{Magnetization Process and Collective Excitations in the $S\mathbf{=}1/2$
					Triangular-Lattice Heisenberg Antiferromagnet
					${\mathrm{Ba}}_{3}{\mathrm{CoSb}}_{2}{\mathbf{O}}_{9}$}},}\ }\href
	{https://doi.org/10.1103/PhysRevLett.110.267201} {\bibfield  {journal}
		{\bibinfo  {journal} {Phys. Rev. Lett.}\ }\textbf {\bibinfo {volume} {110}},\
		\bibinfo {pages} {267201} (\bibinfo {year} {2013})}\BibitemShut {NoStop}%
	\bibitem [{\citenamefont {Dagotto}(1999)}]{Dagotto1999}%
	\BibitemOpen
	\bibfield  {author} {\bibinfo {author} {\bibfnamefont {E.}~\bibnamefont
			{Dagotto}},\ }\bibfield  {title} {\enquote {\bibinfo {title} {Experiments on
				ladders reveal a complex interplay between a spin-gapped normal state and
				superconductivity},}\ }\href {https://doi.org/10.1088/0034-4885/62/11/202}
	{\bibfield  {journal} {\bibinfo  {journal} {Reports on Progress in Physics}\
		}\textbf {\bibinfo {volume} {62}},\ \bibinfo {pages} {1525} (\bibinfo {year}
		{1999})}\BibitemShut {NoStop}%
	\bibitem [{\citenamefont {Wang}\ \emph {et~al.}(2018)\citenamefont {Wang},
		\citenamefont {Wu}, \citenamefont {Yang}, \citenamefont {Bera}, \citenamefont
		{Kamenskyi}, \citenamefont {Islam}, \citenamefont {Xu}, \citenamefont {Law},
		\citenamefont {Lake}, \citenamefont {Wu},\ and\ \citenamefont
		{Loidl}}]{Wang2018_Nature}%
	\BibitemOpen
	\bibfield  {author} {\bibinfo {author} {\bibfnamefont {Z.}~\bibnamefont
			{Wang}}, \bibinfo {author} {\bibfnamefont {J.}~\bibnamefont {Wu}}, \bibinfo
		{author} {\bibfnamefont {W.}~\bibnamefont {Yang}}, \bibinfo {author}
		{\bibfnamefont {A.~K.}\ \bibnamefont {Bera}}, \bibinfo {author}
		{\bibfnamefont {D.}~\bibnamefont {Kamenskyi}}, \bibinfo {author}
		{\bibfnamefont {A.~T. M.~N.}\ \bibnamefont {Islam}}, \bibinfo {author}
		{\bibfnamefont {S.}~\bibnamefont {Xu}}, \bibinfo {author} {\bibfnamefont
			{J.~M.}\ \bibnamefont {Law}}, \bibinfo {author} {\bibfnamefont
			{B.}~\bibnamefont {Lake}}, \bibinfo {author} {\bibfnamefont {C.}~\bibnamefont
			{Wu}},\ and\ \bibinfo {author} {\bibfnamefont {A.}~\bibnamefont {Loidl}},\
	}\bibfield  {title} {\enquote {\bibinfo {title} {Experimental observation of
				bethe strings},}\ }\href {https://doi.org/10.1038/nature25466} {\bibfield
		{journal} {\bibinfo  {journal} {Nature}\ }\textbf {\bibinfo {volume} {554}},\
		\bibinfo {pages} {219--223} (\bibinfo {year} {2018})}\BibitemShut {NoStop}%
	\bibitem [{\citenamefont {Bera}\ \emph {et~al.}(2020)\citenamefont {Bera},
		\citenamefont {Wu}, \citenamefont {Yang}, \citenamefont {Bewley},
		\citenamefont {Boehm}, \citenamefont {Xu}, \citenamefont {Bartkowiak},
		\citenamefont {Prokhnenko}, \citenamefont {Klemke}, \citenamefont {Islam},
		\citenamefont {Law}, \citenamefont {Wang},\ and\ \citenamefont
		{Lake}}]{Bera2020}%
	\BibitemOpen
	\bibfield  {author} {\bibinfo {author} {\bibfnamefont {A.~K.}\ \bibnamefont
			{Bera}}, \bibinfo {author} {\bibfnamefont {J.}~\bibnamefont {Wu}}, \bibinfo
		{author} {\bibfnamefont {W.}~\bibnamefont {Yang}}, \bibinfo {author}
		{\bibfnamefont {R.}~\bibnamefont {Bewley}}, \bibinfo {author} {\bibfnamefont
			{M.}~\bibnamefont {Boehm}}, \bibinfo {author} {\bibfnamefont
			{J.}~\bibnamefont {Xu}}, \bibinfo {author} {\bibfnamefont {M.}~\bibnamefont
			{Bartkowiak}}, \bibinfo {author} {\bibfnamefont {O.}~\bibnamefont
			{Prokhnenko}}, \bibinfo {author} {\bibfnamefont {B.}~\bibnamefont {Klemke}},
		\bibinfo {author} {\bibfnamefont {A.~T. M.~N.}\ \bibnamefont {Islam}},
		\bibinfo {author} {\bibfnamefont {J.~M.}\ \bibnamefont {Law}}, \bibinfo
		{author} {\bibfnamefont {Z.}~\bibnamefont {Wang}},\ and\ \bibinfo {author}
		{\bibfnamefont {B.}~\bibnamefont {Lake}},\ }\bibfield  {title} {\enquote
		{\bibinfo {title} {Dispersions of many-body bethe strings},}\ }\href
	{https://doi.org/10.1038/s41567-020-0835-7} {\bibfield  {journal} {\bibinfo
			{journal} {Nature Physics}\ }\textbf {\bibinfo {volume} {16}},\ \bibinfo
		{pages} {625--630} (\bibinfo {year} {2020})}\BibitemShut {NoStop}%
	\bibitem [{\citenamefont {Faure}\ \emph {et~al.}(2018)\citenamefont {Faure},
		\citenamefont {Takayoshi}, \citenamefont {Petit}, \citenamefont {Simonet},
		\citenamefont {Raymond}, \citenamefont {Regnault}, \citenamefont {Boehm},
		\citenamefont {White}, \citenamefont {M{\aa}nsson}, \citenamefont
		{R{\"u}egg}, \citenamefont {Lejay}, \citenamefont {Canals}, \citenamefont
		{Lorenz}, \citenamefont {Furuya}, \citenamefont {Giamarchi},\ and\
		\citenamefont {Grenier}}]{Faure2018}%
	\BibitemOpen
	\bibfield  {author} {\bibinfo {author} {\bibfnamefont {Q.}~\bibnamefont
			{Faure}}, \bibinfo {author} {\bibfnamefont {S.}~\bibnamefont {Takayoshi}},
		\bibinfo {author} {\bibfnamefont {S.}~\bibnamefont {Petit}}, \bibinfo
		{author} {\bibfnamefont {V.}~\bibnamefont {Simonet}}, \bibinfo {author}
		{\bibfnamefont {S.}~\bibnamefont {Raymond}}, \bibinfo {author} {\bibfnamefont
			{L.-P.}\ \bibnamefont {Regnault}}, \bibinfo {author} {\bibfnamefont
			{M.}~\bibnamefont {Boehm}}, \bibinfo {author} {\bibfnamefont {J.~S.}\
			\bibnamefont {White}}, \bibinfo {author} {\bibfnamefont {M.}~\bibnamefont
			{M{\aa}nsson}}, \bibinfo {author} {\bibfnamefont {C.}~\bibnamefont
			{R{\"u}egg}}, \bibinfo {author} {\bibfnamefont {P.}~\bibnamefont {Lejay}},
		\bibinfo {author} {\bibfnamefont {B.}~\bibnamefont {Canals}}, \bibinfo
		{author} {\bibfnamefont {T.}~\bibnamefont {Lorenz}}, \bibinfo {author}
		{\bibfnamefont {S.~C.}\ \bibnamefont {Furuya}}, \bibinfo {author}
		{\bibfnamefont {T.}~\bibnamefont {Giamarchi}},\ and\ \bibinfo {author}
		{\bibfnamefont {B.}~\bibnamefont {Grenier}},\ }\bibfield  {title} {\enquote
		{\bibinfo {title} {{Topological quantum phase transition in the Ising-like
					antiferromagnetic spin chain BaCo$_2$V$_2$O$_8$}},}\ }\href
	{https://doi.org/10.1038/s41567-018-0126-8} {\bibfield  {journal} {\bibinfo
			{journal} {Nature Physics}\ }\textbf {\bibinfo {volume} {14}},\ \bibinfo
		{pages} {716--722} (\bibinfo {year} {2018})}\BibitemShut {NoStop}%
	\bibitem [{\citenamefont {Liu}\ and\ \citenamefont
		{Khaliullin}(2018)}]{Liu2018}%
	\BibitemOpen
	\bibfield  {author} {\bibinfo {author} {\bibfnamefont {H.}~\bibnamefont
			{Liu}}\ and\ \bibinfo {author} {\bibfnamefont {G.}~\bibnamefont
			{Khaliullin}},\ }\bibfield  {title} {\enquote {\bibinfo {title} {Pseudospin
				exchange interactions in ${d}^{7}$ cobalt compounds: Possible realization of
				the kitaev model},}\ }\href {https://doi.org/10.1103/PhysRevB.97.014407}
	{\bibfield  {journal} {\bibinfo  {journal} {Phys. Rev. B}\ }\textbf {\bibinfo
			{volume} {97}},\ \bibinfo {pages} {014407} (\bibinfo {year}
		{2018})}\BibitemShut {NoStop}%
	\bibitem [{\citenamefont {Liu}, \citenamefont {Chaloupka},\ and\ \citenamefont
		{Khaliullin}(2020)}]{Liu2020}%
	\BibitemOpen
	\bibfield  {author} {\bibinfo {author} {\bibfnamefont {H.}~\bibnamefont
			{Liu}}, \bibinfo {author} {\bibfnamefont {J.}~\bibnamefont {Chaloupka}},\
		and\ \bibinfo {author} {\bibfnamefont {G.}~\bibnamefont {Khaliullin}},\
	}\bibfield  {title} {\enquote {\bibinfo {title} {Kitaev spin liquid in $3d$
				transition metal compounds},}\ }\href
	{https://doi.org/10.1103/PhysRevLett.125.047201} {\bibfield  {journal}
		{\bibinfo  {journal} {Phys. Rev. Lett.}\ }\textbf {\bibinfo {volume} {125}},\
		\bibinfo {pages} {047201} (\bibinfo {year} {2020})}\BibitemShut {NoStop}%
	\bibitem [{\citenamefont {Xiang}\ \emph {et~al.}(2024)\citenamefont {Xiang},
		\citenamefont {Zhang}, \citenamefont {Gao}, \citenamefont {Schmidt},
		\citenamefont {Schmalzl}, \citenamefont {Wang}, \citenamefont {Li},
		\citenamefont {Xi}, \citenamefont {Liu}, \citenamefont {Jin}, \citenamefont
		{Li}, \citenamefont {Shen}, \citenamefont {Chen}, \citenamefont {Qi},
		\citenamefont {Wan}, \citenamefont {Jin}, \citenamefont {Li}, \citenamefont
		{Sun},\ and\ \citenamefont {Su}}]{Xiang2024}%
	\BibitemOpen
	\bibfield  {author} {\bibinfo {author} {\bibfnamefont {J.}~\bibnamefont
			{Xiang}}, \bibinfo {author} {\bibfnamefont {C.}~\bibnamefont {Zhang}},
		\bibinfo {author} {\bibfnamefont {Y.}~\bibnamefont {Gao}}, \bibinfo {author}
		{\bibfnamefont {W.}~\bibnamefont {Schmidt}}, \bibinfo {author} {\bibfnamefont
			{K.}~\bibnamefont {Schmalzl}}, \bibinfo {author} {\bibfnamefont {C.-W.}\
			\bibnamefont {Wang}}, \bibinfo {author} {\bibfnamefont {B.}~\bibnamefont
			{Li}}, \bibinfo {author} {\bibfnamefont {N.}~\bibnamefont {Xi}}, \bibinfo
		{author} {\bibfnamefont {X.-Y.}\ \bibnamefont {Liu}}, \bibinfo {author}
		{\bibfnamefont {H.}~\bibnamefont {Jin}}, \bibinfo {author} {\bibfnamefont
			{G.}~\bibnamefont {Li}}, \bibinfo {author} {\bibfnamefont {J.}~\bibnamefont
			{Shen}}, \bibinfo {author} {\bibfnamefont {Z.}~\bibnamefont {Chen}}, \bibinfo
		{author} {\bibfnamefont {Y.}~\bibnamefont {Qi}}, \bibinfo {author}
		{\bibfnamefont {Y.}~\bibnamefont {Wan}}, \bibinfo {author} {\bibfnamefont
			{W.}~\bibnamefont {Jin}}, \bibinfo {author} {\bibfnamefont {W.}~\bibnamefont
			{Li}}, \bibinfo {author} {\bibfnamefont {P.}~\bibnamefont {Sun}},\ and\
		\bibinfo {author} {\bibfnamefont {G.}~\bibnamefont {Su}},\ }\bibfield
	{title} {\enquote {\bibinfo {title} {{Giant magnetocaloric effect in spin
					supersolid candidate Na$_2$BaCo(PO$_4$)$_2$}},}\ }\href
	{https://doi.org/10.1038/s41586-023-06885-w} {\bibfield  {journal} {\bibinfo
			{journal} {Nature}\ }\textbf {\bibinfo {volume} {625}},\ \bibinfo {pages}
		{270--275} (\bibinfo {year} {2024})}\BibitemShut {NoStop}%
	\bibitem [{\citenamefont {Chen}(2024)}]{Chen2024}%
	\BibitemOpen
	\bibfield  {author} {\bibinfo {author} {\bibfnamefont {G.~V.}\ \bibnamefont
			{Chen}},\ }\bibfield  {title} {\enquote {\bibinfo {title} {{Emergent
					Berezinskii-Kosterlitz-Thouless and Kugel-Khomskii Physics in the Triangular
					Lattice Bilayer Colbaltate}},}\ }\href
	{https://doi.org/10.1103/PhysRevLett.133.136703} {\bibfield  {journal}
		{\bibinfo  {journal} {Phys. Rev. Lett.}\ }\textbf {\bibinfo {volume} {133}},\
		\bibinfo {pages} {136703} (\bibinfo {year} {2024})}\BibitemShut {NoStop}%
	\bibitem [{\citenamefont {Zhu}, \citenamefont {Gong},\ and\ \citenamefont
		{Sheng}(2019)}]{Zhu2019}%
	\BibitemOpen
	\bibfield  {author} {\bibinfo {author} {\bibfnamefont {W.}~\bibnamefont
			{Zhu}}, \bibinfo {author} {\bibfnamefont {S.~S.}\ \bibnamefont {Gong}},\ and\
		\bibinfo {author} {\bibfnamefont {D.~N.}\ \bibnamefont {Sheng}},\ }\bibfield
	{title} {\enquote {\bibinfo {title} {{Competing Phases in Spin-1/2 Heisenberg
					Model on a Honeycomb Lattice with Next-Nearest-Neighbor Coupling}},}\ }\href
	{https://doi.org/10.1103/PhysRevB.100.184422} {\bibfield  {journal} {\bibinfo
			{journal} {Physical Review B}\ }\textbf {\bibinfo {volume} {100}},\ \bibinfo
		{pages} {184422} (\bibinfo {year} {2019})}\BibitemShut {NoStop}%
	\bibitem [{\citenamefont {Liu}, \citenamefont {Tseng},\ and\ \citenamefont
		{Kao}(2019)}]{Liu2019}%
	\BibitemOpen
	\bibfield  {author} {\bibinfo {author} {\bibfnamefont {C.~W.}\ \bibnamefont
			{Liu}}, \bibinfo {author} {\bibfnamefont {C.~Y.}\ \bibnamefont {Tseng}},\
		and\ \bibinfo {author} {\bibfnamefont {Y.~J.}\ \bibnamefont {Kao}},\
	}\bibfield  {title} {\enquote {\bibinfo {title} {{Quantum Phase Transition in
					the Spin-1/2 Antiferromagnetic Heisenberg Chain with Competing
					Interactions}},}\ }\href {https://doi.org/10.1103/PhysRevLett.122.017201}
	{\bibfield  {journal} {\bibinfo  {journal} {Physical Review Letters}\
		}\textbf {\bibinfo {volume} {122}},\ \bibinfo {pages} {017201} (\bibinfo
		{year} {2019})}\BibitemShut {NoStop}%
	\bibitem [{\citenamefont {Sato}\ \emph {et~al.}(2011)\citenamefont {Sato},
		\citenamefont {Furukawa}, \citenamefont {Onoda},\ and\ \citenamefont
		{Furusaki}}]{Sato2011}%
	\BibitemOpen
	\bibfield  {author} {\bibinfo {author} {\bibfnamefont {M.}~\bibnamefont
			{Sato}}, \bibinfo {author} {\bibfnamefont {S.}~\bibnamefont {Furukawa}},
		\bibinfo {author} {\bibfnamefont {S.}~\bibnamefont {Onoda}},\ and\ \bibinfo
		{author} {\bibfnamefont {A.}~\bibnamefont {Furusaki}},\ }\bibfield  {title}
	{\enquote {\bibinfo {title} {Competing phases in spin-1/2 {J1-J2} chain with
				easy-plane anisotropy},}\ }\href {https://doi.org/10.1142/S0217984911026607}
	{\bibfield  {journal} {\bibinfo  {journal} {Modern Physics Letters B}\
		}\textbf {\bibinfo {volume} {25}},\ \bibinfo {pages} {901--908} (\bibinfo
		{year} {2011})}\BibitemShut {NoStop}%
	\bibitem [{\citenamefont {Hirobe}\ \emph {et~al.}(2017)\citenamefont {Hirobe},
		\citenamefont {Sato}, \citenamefont {Kawamata}, \citenamefont {Shiomi},
		\citenamefont {Uchida}, \citenamefont {Iguchi}, \citenamefont {Koike},
		\citenamefont {Maekawa},\ and\ \citenamefont {Saitoh}}]{Hirobe2017}%
	\BibitemOpen
	\bibfield  {author} {\bibinfo {author} {\bibfnamefont {D.}~\bibnamefont
			{Hirobe}}, \bibinfo {author} {\bibfnamefont {M.}~\bibnamefont {Sato}},
		\bibinfo {author} {\bibfnamefont {T.}~\bibnamefont {Kawamata}}, \bibinfo
		{author} {\bibfnamefont {Y.}~\bibnamefont {Shiomi}}, \bibinfo {author}
		{\bibfnamefont {K.-i.}\ \bibnamefont {Uchida}}, \bibinfo {author}
		{\bibfnamefont {R.}~\bibnamefont {Iguchi}}, \bibinfo {author} {\bibfnamefont
			{Y.}~\bibnamefont {Koike}}, \bibinfo {author} {\bibfnamefont
			{S.}~\bibnamefont {Maekawa}},\ and\ \bibinfo {author} {\bibfnamefont
			{E.}~\bibnamefont {Saitoh}},\ }\bibfield  {title} {\enquote {\bibinfo {title}
			{One-dimensional spinon spin currents},}\ }\href
	{https://doi.org/10.1038/nphys3895} {\bibfield  {journal} {\bibinfo
			{journal} {Nature Physics}\ }\textbf {\bibinfo {volume} {13}},\ \bibinfo
		{pages} {30--34} (\bibinfo {year} {2017})}\BibitemShut {NoStop}%
	\bibitem [{\citenamefont {Li}\ and\ \citenamefont {Comin}(2023)}]{Li2024}%
	\BibitemOpen
	\bibfield  {author} {\bibinfo {author} {\bibfnamefont {J.}~\bibnamefont
			{Li}}\ and\ \bibinfo {author} {\bibfnamefont {R.}~\bibnamefont {Comin}},\
	}\bibfield  {title} {\enquote {\bibinfo {title} {The first free-standing 1d
				spin chain},}\ }\href
	{https://doi.org/https://doi.org/10.1016/j.matt.2023.06.025} {\bibfield
		{journal} {\bibinfo  {journal} {Matter}\ }\textbf {\bibinfo {volume} {6}},\
		\bibinfo {pages} {2576--2578} (\bibinfo {year} {2023})}\BibitemShut {NoStop}%
	\bibitem [{\citenamefont {Yang}\ and\ \citenamefont {Yang}(1966)}]{Yang1966}%
	\BibitemOpen
	\bibfield  {author} {\bibinfo {author} {\bibfnamefont {C.~N.}\ \bibnamefont
			{Yang}}\ and\ \bibinfo {author} {\bibfnamefont {C.~P.}\ \bibnamefont
			{Yang}},\ }\bibfield  {title} {\enquote {\bibinfo {title} {{One-Dimensional
					Chain of Anisotropic Spin-Spin Interactions. I. Proof of Bethe's Hypothesis
					for Ground State in a Finite System}},}\ }\href
	{https://doi.org/10.1103/PhysRev.150.321} {\bibfield  {journal} {\bibinfo
			{journal} {Phys. Rev.}\ }\textbf {\bibinfo {volume} {150}},\ \bibinfo {pages}
		{321--327} (\bibinfo {year} {1966})}\BibitemShut {NoStop}%
	\bibitem [{\citenamefont {Isha}\ \emph {et~al.}(2024)\citenamefont {Isha},
		\citenamefont {Bera}, \citenamefont {Guratinder}, \citenamefont {Stock},
		\citenamefont {Chakraborty}, \citenamefont {Puphal}, \citenamefont {Isobe},
		\citenamefont {K\"uster}, \citenamefont {Skourski}, \citenamefont
		{Bhaskaran}, \citenamefont {Zvyagin}, \citenamefont {Luther}, \citenamefont
		{Gronemann}, \citenamefont {K\"uhne}, \citenamefont {Salazar~Mej\'{\i}a},
		\citenamefont {Pregelj}, \citenamefont {Hansen}, \citenamefont {Kaushik},
		\citenamefont {Voneshen}, \citenamefont {Kulkarni}, \citenamefont {Lalla},
		\citenamefont {Yusuf}, \citenamefont {Thamizhavel},\ and\ \citenamefont
		{Yogi}}]{Isha2024}%
	\BibitemOpen
	\bibfield  {author} {\bibinfo {author} {\bibnamefont {Isha}}, \bibinfo
		{author} {\bibfnamefont {A.~K.}\ \bibnamefont {Bera}}, \bibinfo {author}
		{\bibfnamefont {K.}~\bibnamefont {Guratinder}}, \bibinfo {author}
		{\bibfnamefont {C.}~\bibnamefont {Stock}}, \bibinfo {author} {\bibfnamefont
			{K.}~\bibnamefont {Chakraborty}}, \bibinfo {author} {\bibfnamefont
			{P.}~\bibnamefont {Puphal}}, \bibinfo {author} {\bibfnamefont
			{M.}~\bibnamefont {Isobe}}, \bibinfo {author} {\bibfnamefont
			{K.}~\bibnamefont {K\"uster}}, \bibinfo {author} {\bibfnamefont
			{Y.}~\bibnamefont {Skourski}}, \bibinfo {author} {\bibfnamefont
			{L.}~\bibnamefont {Bhaskaran}}, \bibinfo {author} {\bibfnamefont {S.~A.}\
			\bibnamefont {Zvyagin}}, \bibinfo {author} {\bibfnamefont {S.}~\bibnamefont
			{Luther}}, \bibinfo {author} {\bibfnamefont {J.}~\bibnamefont {Gronemann}},
		\bibinfo {author} {\bibfnamefont {H.}~\bibnamefont {K\"uhne}}, \bibinfo
		{author} {\bibfnamefont {C.}~\bibnamefont {Salazar~Mej\'{\i}a}}, \bibinfo
		{author} {\bibfnamefont {M.}~\bibnamefont {Pregelj}}, \bibinfo {author}
		{\bibfnamefont {T.~C.}\ \bibnamefont {Hansen}}, \bibinfo {author}
		{\bibfnamefont {S.~D.}\ \bibnamefont {Kaushik}}, \bibinfo {author}
		{\bibfnamefont {D.}~\bibnamefont {Voneshen}}, \bibinfo {author}
		{\bibfnamefont {R.}~\bibnamefont {Kulkarni}}, \bibinfo {author}
		{\bibfnamefont {N.~P.}\ \bibnamefont {Lalla}}, \bibinfo {author}
		{\bibfnamefont {S.~M.}\ \bibnamefont {Yusuf}}, \bibinfo {author}
		{\bibfnamefont {A.}~\bibnamefont {Thamizhavel}},\ and\ \bibinfo {author}
		{\bibfnamefont {A.~K.}\ \bibnamefont {Yogi}},\ }\bibfield  {title} {\enquote
		{\bibinfo {title} {{Sharp quantum phase transition in the frustrated spin-1/2
					Ising chain antiferromagnet ${\mathrm{CaCoV}}_{2}{\mathrm{O}}_{7}$}},}\
	}\href {https://doi.org/10.1103/PhysRevResearch.6.L032010} {\bibfield
		{journal} {\bibinfo  {journal} {Phys. Rev. Res.}\ }\textbf {\bibinfo {volume}
			{6}},\ \bibinfo {pages} {L032010} (\bibinfo {year} {2024})}\BibitemShut
	{NoStop}%
	\bibitem [{\citenamefont {Petříček}, \citenamefont {Dušek},\ and\
		\citenamefont {Palatinus}(2014)}]{Jana2006}%
	\BibitemOpen
	\bibfield  {author} {\bibinfo {author} {\bibfnamefont {V.}~\bibnamefont
			{Petříček}}, \bibinfo {author} {\bibfnamefont {M.}~\bibnamefont
			{Dušek}},\ and\ \bibinfo {author} {\bibfnamefont {L.}~\bibnamefont
			{Palatinus}},\ }\bibfield  {title} {\enquote {\bibinfo {title}
			{{Crystallographic Computing System JANA2006: General features}},}\ }\href
	{https://doi.org/doi:10.1515/zkri-2014-1737} {\bibfield  {journal} {\bibinfo
			{journal} {Zeitschrift für Kristallographie - Crystalline Materials}\
		}\textbf {\bibinfo {volume} {229}},\ \bibinfo {pages} {345--352} (\bibinfo
		{year} {2014})}\BibitemShut {NoStop}%
	\bibitem [{\citenamefont {Ouladdiaf}\ \emph {et~al.}(2006)\citenamefont
		{Ouladdiaf}, \citenamefont {Archer}, \citenamefont {McIntyre}, \citenamefont
		{Hewat}, \citenamefont {Brau},\ and\ \citenamefont {York}}]{Laue}%
	\BibitemOpen
	\bibfield  {author} {\bibinfo {author} {\bibfnamefont {B.}~\bibnamefont
			{Ouladdiaf}}, \bibinfo {author} {\bibfnamefont {J.}~\bibnamefont {Archer}},
		\bibinfo {author} {\bibfnamefont {G.}~\bibnamefont {McIntyre}}, \bibinfo
		{author} {\bibfnamefont {A.}~\bibnamefont {Hewat}}, \bibinfo {author}
		{\bibfnamefont {D.}~\bibnamefont {Brau}},\ and\ \bibinfo {author}
		{\bibfnamefont {S.}~\bibnamefont {York}},\ }\bibfield  {title} {\enquote
		{\bibinfo {title} {{OrientExpress: A new system for Laue neutron
					diffraction}},}\ }\href
	{https://doi.org/https://doi.org/10.1016/j.physb.2006.05.337} {\bibfield
		{journal} {\bibinfo  {journal} {Physica B: Condensed Matter}\ }\textbf
		{\bibinfo {volume} {385-386}},\ \bibinfo {pages} {1052--1054} (\bibinfo
		{year} {2006})}\BibitemShut {NoStop}%
	\bibitem [{\citenamefont {Murashova}, \citenamefont {Velikodnyi},\ and\
		\citenamefont {Zhuravlev}(1994)}]{Murashova1994}%
	\BibitemOpen
	\bibfield  {author} {\bibinfo {author} {\bibfnamefont {E.~V.}\ \bibnamefont
			{Murashova}}, \bibinfo {author} {\bibfnamefont {Y.~A.}\ \bibnamefont
			{Velikodnyi}},\ and\ \bibinfo {author} {\bibfnamefont {V.~D.}\ \bibnamefont
			{Zhuravlev}},\ }\bibfield  {title} {\enquote {\bibinfo {title} {{ChemInform
					Abstract: Crystal Structure of the Double Pyrovanadates {CaMgV$_2$O7} and
					{CaCoV$_2$O$_7$}.}}}\ }\href
	{https://doi.org/https://doi.org/10.1002/chin.199413007} {\bibfield
		{journal} {\bibinfo  {journal} {ChemInform}\ }\textbf {\bibinfo {volume}
			{25}} (\bibinfo {year} {1994}),\
		https://doi.org/10.1002/chin.199413007}\BibitemShut {NoStop}%
	\bibitem [{\citenamefont {Murasaki}\ \emph {et~al.}(2021)\citenamefont
		{Murasaki}, \citenamefont {Nawa}, \citenamefont {Okuyama}, \citenamefont
		{Avdeev},\ and\ \citenamefont {Sato}}]{Murasaki2021}%
	\BibitemOpen
	\bibfield  {author} {\bibinfo {author} {\bibfnamefont {R.}~\bibnamefont
			{Murasaki}}, \bibinfo {author} {\bibfnamefont {K.}~\bibnamefont {Nawa}},
		\bibinfo {author} {\bibfnamefont {D.}~\bibnamefont {Okuyama}}, \bibinfo
		{author} {\bibfnamefont {M.}~\bibnamefont {Avdeev}},\ and\ \bibinfo {author}
		{\bibfnamefont {T.~J.}\ \bibnamefont {Sato}},\ }\bibfield  {title} {\enquote
		{\bibinfo {title} {Antiferromagnetic order of ferromagnetically coupled
				dimers in the double pyrovanadate {CaCoV$_2$O$_7$}},}\ }\href@noop {} {\
		(\bibinfo {year} {2021})},\ \Eprint {https://arxiv.org/abs/2108.00715}
	{2108.00715 [cond-mat.mtrl-sci]} \BibitemShut {NoStop}%
	\bibitem [{\citenamefont {Zhuravlev}\ \emph {et~al.}(2018)\citenamefont
		{Zhuravlev}, \citenamefont {Tyutyunnik}, \citenamefont {Chufarov},
		\citenamefont {Lobachevskaya},\ and\ \citenamefont
		{Velikodnyi}}]{Zhuravlev2018}%
	\BibitemOpen
	\bibfield  {author} {\bibinfo {author} {\bibfnamefont {V.}~\bibnamefont
			{Zhuravlev}}, \bibinfo {author} {\bibfnamefont {A.}~\bibnamefont
			{Tyutyunnik}}, \bibinfo {author} {\bibfnamefont {A.}~\bibnamefont
			{Chufarov}}, \bibinfo {author} {\bibfnamefont {N.}~\bibnamefont
			{Lobachevskaya}},\ and\ \bibinfo {author} {\bibfnamefont {A.}~\bibnamefont
			{Velikodnyi}},\ }\bibfield  {title} {\enquote {\bibinfo {title} {{Crystal
					structure of Ca$_2$Zn$_2$ (V$_4$O$_{14}$) and Pb$_2$Cd$_2$
					(V$_3$O$_{10}$)(VO$_4$) double vanadates}},}\ }\href
	{https://doi.org/10.1017/S0885715618000441} {\bibfield  {journal} {\bibinfo
			{journal} {Powder Diffraction}\ }\textbf {\bibinfo {volume} {33}},\ \bibinfo
		{pages} {216--224} (\bibinfo {year} {2018})}\BibitemShut {NoStop}%
	\bibitem [{\citenamefont {Babaryk}\ \emph {et~al.}(2015)\citenamefont
		{Babaryk}, \citenamefont {Odynets}, \citenamefont {Khainakov}, \citenamefont
		{Garcia-Granda},\ and\ \citenamefont {Slobodyanik}}]{Babaryk2015}%
	\BibitemOpen
	\bibfield  {author} {\bibinfo {author} {\bibfnamefont {A.~A.}\ \bibnamefont
			{Babaryk}}, \bibinfo {author} {\bibfnamefont {I.~V.}\ \bibnamefont
			{Odynets}}, \bibinfo {author} {\bibfnamefont {S.}~\bibnamefont {Khainakov}},
		\bibinfo {author} {\bibfnamefont {S.}~\bibnamefont {Garcia-Granda}},\ and\
		\bibinfo {author} {\bibfnamefont {N.~S.}\ \bibnamefont {Slobodyanik}},\
	}\bibfield  {title} {\enquote {\bibinfo {title} {{Polyanionic identity of
					Ca$_2$Zn$_2$(V$_3$O$_{10}$)(VO$_4$) photocatalyst manifested by X-ray powder
					diffraction and periodic boundary density functional theory calculations}},}\
	}\href {https://doi.org/10.1039/C5CE01212K} {\bibfield  {journal} {\bibinfo
			{journal} {CrystEngComm}\ }\textbf {\bibinfo {volume} {17}},\ \bibinfo
		{pages} {7772--7777} (\bibinfo {year} {2015})}\BibitemShut {NoStop}%
	\bibitem [{\citenamefont {Huang}\ \emph {et~al.}(2022)\citenamefont {Huang},
		\citenamefont {Huang}, \citenamefont {Hsu},\ and\ \citenamefont
		{Huang}}]{Huang2022}%
	\BibitemOpen
	\bibfield  {author} {\bibinfo {author} {\bibfnamefont {Y.-T.}\ \bibnamefont
			{Huang}}, \bibinfo {author} {\bibfnamefont {C.-C.}\ \bibnamefont {Huang}},
		\bibinfo {author} {\bibfnamefont {T.-H.}\ \bibnamefont {Hsu}},\ and\ \bibinfo
		{author} {\bibfnamefont {C.-L.}\ \bibnamefont {Huang}},\ }\bibfield  {title}
	{\enquote {\bibinfo {title} {{Ultra-low temperature sintering and microwave
					dielectric properties of Mg-substituted {SrCoV$_2$O$_7$} ceramics}},}\ }\href
	{https://doi.org/10.1080/21870764.2022.2031535} {\bibfield  {journal}
		{\bibinfo  {journal} {Journal of Asian Ceramic Societies}\ }\textbf {\bibinfo
			{volume} {10}},\ \bibinfo {pages} {188--195} (\bibinfo {year}
		{2022})}\BibitemShut {NoStop}%
	\bibitem [{\citenamefont {Greczynski}(2024)}]{Grzegorz2024}%
	\BibitemOpen
	\bibfield  {author} {\bibinfo {author} {\bibfnamefont {G.}~\bibnamefont
			{Greczynski}},\ }\bibfield  {title} {\enquote {\bibinfo {title} {{Binding
					energy referencing in X-ray photoelectron spectroscopy: Expanded data set
					confirms that adventitious carbon aligns to the sample vacuum level}},}\
	}\href {https://doi.org/https://doi.org/10.1016/j.apsusc.2024.160666}
	{\bibfield  {journal} {\bibinfo  {journal} {Applied Surface Science}\
		}\textbf {\bibinfo {volume} {670}},\ \bibinfo {pages} {160666} (\bibinfo
		{year} {2024})}\BibitemShut {NoStop}%
	\bibitem [{\citenamefont {Takubo}\ \emph {et~al.}(2005)\citenamefont {Takubo},
		\citenamefont {Mizokawa}, \citenamefont {Hirata}, \citenamefont {Son},
		\citenamefont {Fujimori}, \citenamefont {Topwal}, \citenamefont {Sarma},
		\citenamefont {Rayaprol},\ and\ \citenamefont {Sampathkumaran}}]{Takubo2005}%
	\BibitemOpen
	\bibfield  {author} {\bibinfo {author} {\bibfnamefont {K.}~\bibnamefont
			{Takubo}}, \bibinfo {author} {\bibfnamefont {T.}~\bibnamefont {Mizokawa}},
		\bibinfo {author} {\bibfnamefont {S.}~\bibnamefont {Hirata}}, \bibinfo
		{author} {\bibfnamefont {J.-Y.}\ \bibnamefont {Son}}, \bibinfo {author}
		{\bibfnamefont {A.}~\bibnamefont {Fujimori}}, \bibinfo {author}
		{\bibfnamefont {D.}~\bibnamefont {Topwal}}, \bibinfo {author} {\bibfnamefont
			{D.~D.}\ \bibnamefont {Sarma}}, \bibinfo {author} {\bibfnamefont
			{S.}~\bibnamefont {Rayaprol}},\ and\ \bibinfo {author} {\bibfnamefont
			{E.-V.}\ \bibnamefont {Sampathkumaran}},\ }\bibfield  {title} {\enquote
		{\bibinfo {title} {{Electronic structure of
					${\mathrm{Ca}}_{3}\mathrm{Co}X{\mathrm{O}}_{6}$ ($X=\mathrm{Co}$, Rh, Ir)
					studied by x-ray photoemission spectroscopy}},}\ }\href
	{https://doi.org/10.1103/PhysRevB.71.073406} {\bibfield  {journal} {\bibinfo
			{journal} {Phys. Rev. B}\ }\textbf {\bibinfo {volume} {71}},\ \bibinfo
		{pages} {073406} (\bibinfo {year} {2005})}\BibitemShut {NoStop}%
	\bibitem [{\citenamefont {van Elp}\ \emph {et~al.}(1991)\citenamefont {van
			Elp}, \citenamefont {Wieland}, \citenamefont {Eskes}, \citenamefont {Kuiper},
		\citenamefont {Sawatzky}, \citenamefont {de~Groot},\ and\ \citenamefont
		{Turner}}]{Elp1991}%
	\BibitemOpen
	\bibfield  {author} {\bibinfo {author} {\bibfnamefont {J.}~\bibnamefont {van
				Elp}}, \bibinfo {author} {\bibfnamefont {J.~L.}\ \bibnamefont {Wieland}},
		\bibinfo {author} {\bibfnamefont {H.}~\bibnamefont {Eskes}}, \bibinfo
		{author} {\bibfnamefont {P.}~\bibnamefont {Kuiper}}, \bibinfo {author}
		{\bibfnamefont {G.~A.}\ \bibnamefont {Sawatzky}}, \bibinfo {author}
		{\bibfnamefont {F.~M.~F.}\ \bibnamefont {de~Groot}},\ and\ \bibinfo {author}
		{\bibfnamefont {T.~S.}\ \bibnamefont {Turner}},\ }\bibfield  {title}
	{\enquote {\bibinfo {title} {{Electronic structure of CoO, Li-doped CoO, and
					${\mathrm{LiCoO}}_{2}$}},}\ }\href {https://doi.org/10.1103/PhysRevB.44.6090}
	{\bibfield  {journal} {\bibinfo  {journal} {Phys. Rev. B}\ }\textbf {\bibinfo
			{volume} {44}},\ \bibinfo {pages} {6090--6103} (\bibinfo {year}
		{1991})}\BibitemShut {NoStop}%
	\bibitem [{\citenamefont {de~Groot}\ \emph {et~al.}(1993)\citenamefont
		{de~Groot}, \citenamefont {Abbate}, \citenamefont {van Elp}, \citenamefont
		{Sawatzky}, \citenamefont {Ma}, \citenamefont {Chen},\ and\ \citenamefont
		{Sette}}]{Groot1993}%
	\BibitemOpen
	\bibfield  {author} {\bibinfo {author} {\bibfnamefont {F.~M.~F.}\
			\bibnamefont {de~Groot}}, \bibinfo {author} {\bibfnamefont {M.}~\bibnamefont
			{Abbate}}, \bibinfo {author} {\bibfnamefont {J.}~\bibnamefont {van Elp}},
		\bibinfo {author} {\bibfnamefont {G.~A.}\ \bibnamefont {Sawatzky}}, \bibinfo
		{author} {\bibfnamefont {Y.~J.}\ \bibnamefont {Ma}}, \bibinfo {author}
		{\bibfnamefont {C.~T.}\ \bibnamefont {Chen}},\ and\ \bibinfo {author}
		{\bibfnamefont {F.}~\bibnamefont {Sette}},\ }\bibfield  {title} {\enquote
		{\bibinfo {title} {{Oxygen 1s and cobalt 2p X-ray absorption of cobalt
					oxides}},}\ }\href {https://doi.org/10.1088/0953-8984/5/14/023} {\bibfield
		{journal} {\bibinfo  {journal} {Journal of Physics: Condensed Matter}\
		}\textbf {\bibinfo {volume} {5}},\ \bibinfo {pages} {2277} (\bibinfo {year}
		{1993})}\BibitemShut {NoStop}%
	\bibitem [{\citenamefont {Liardet}\ and\ \citenamefont
		{Hu}(2018)}]{Liardet2018}%
	\BibitemOpen
	\bibfield  {author} {\bibinfo {author} {\bibfnamefont {L.}~\bibnamefont
			{Liardet}}\ and\ \bibinfo {author} {\bibfnamefont {X.}~\bibnamefont {Hu}},\
	}\bibfield  {title} {\enquote {\bibinfo {title} {{Amorphous Cobalt Vanadium
					Oxide as a Highly Active Electrocatalyst for Oxygen Evolution}},}\ }\href
	{https://doi.org/10.1021/acscatal.7b03198} {\bibfield  {journal} {\bibinfo
			{journal} {ACS Catalysis}\ }\textbf {\bibinfo {volume} {8}},\ \bibinfo
		{pages} {644--650} (\bibinfo {year} {2018})},\ \bibinfo {note} {pMID:
		29333330}\BibitemShut {NoStop}%
	\bibitem [{\citenamefont {Alov}\ \emph {et~al.}(2006)\citenamefont {Alov},
		\citenamefont {Kutsko}, \citenamefont {Spirovová},\ and\ \citenamefont
		{Bastl}}]{Alov2006}%
	\BibitemOpen
	\bibfield  {author} {\bibinfo {author} {\bibfnamefont {N.}~\bibnamefont
			{Alov}}, \bibinfo {author} {\bibfnamefont {D.}~\bibnamefont {Kutsko}},
		\bibinfo {author} {\bibfnamefont {I.}~\bibnamefont {Spirovová}},\ and\
		\bibinfo {author} {\bibfnamefont {Z.}~\bibnamefont {Bastl}},\ }\bibfield
	{title} {\enquote {\bibinfo {title} {{XPS study of vanadium surface oxidation
					by oxygen ion bombardment}},}\ }\href
	{https://doi.org/https://doi.org/10.1016/j.susc.2005.12.052} {\bibfield
		{journal} {\bibinfo  {journal} {Surface Science}\ }\textbf {\bibinfo {volume}
			{600}},\ \bibinfo {pages} {1628--1631} (\bibinfo {year} {2006})},\ \bibinfo
	{note} {surface Science}\BibitemShut {NoStop}%
	\bibitem [{\citenamefont {Young}\ and\ \citenamefont
		{Otagawa}(1985)}]{Young1985}%
	\BibitemOpen
	\bibfield  {author} {\bibinfo {author} {\bibfnamefont {V.}~\bibnamefont
			{Young}}\ and\ \bibinfo {author} {\bibfnamefont {T.}~\bibnamefont
			{Otagawa}},\ }\bibfield  {title} {\enquote {\bibinfo {title} {{XPS studies on
					strontium compounds}},}\ }\href
	{https://doi.org/https://doi.org/10.1016/0378-5963(85)90083-2} {\bibfield
		{journal} {\bibinfo  {journal} {Applications of Surface Science}\ }\textbf
		{\bibinfo {volume} {20}},\ \bibinfo {pages} {228--248} (\bibinfo {year}
		{1985})}\BibitemShut {NoStop}%
	\bibitem [{\citenamefont {Mugiraneza}\ and\ \citenamefont
		{Hallas}(2022)}]{Mugiraneza2022}%
	\BibitemOpen
	\bibfield  {author} {\bibinfo {author} {\bibfnamefont {S.}~\bibnamefont
			{Mugiraneza}}\ and\ \bibinfo {author} {\bibfnamefont {A.~M.}\ \bibnamefont
			{Hallas}},\ }\bibfield  {title} {\enquote {\bibinfo {title} {{Tutorial: a
					beginner's guide to interpreting magnetic susceptibility data with the
					Curie-Weiss law}},}\ }\href {https://doi.org/10.1038/s42005-022-00853-y}
	{\bibfield  {journal} {\bibinfo  {journal} {Communications Physics}\ }\textbf
		{\bibinfo {volume} {5}},\ \bibinfo {pages} {95} (\bibinfo {year}
		{2022})}\BibitemShut {NoStop}%
	\bibitem [{\citenamefont {Lin}\ \emph {et~al.}(2021)\citenamefont {Lin},
		\citenamefont {Jeong}, \citenamefont {Kim}, \citenamefont {Wang},
		\citenamefont {Huang}, \citenamefont {Masuda}, \citenamefont {Asai},
		\citenamefont {Itoh}, \citenamefont {Günther}, \citenamefont {Russina},
		\citenamefont {Lu}, \citenamefont {Sheng}, \citenamefont {Wang},
		\citenamefont {Wang}, \citenamefont {Wang}, \citenamefont {Ren},
		\citenamefont {Xi}, \citenamefont {Tong}, \citenamefont {Ling}, \citenamefont
		{Liu}, \citenamefont {Wu}, \citenamefont {Mei}, \citenamefont {Qu},
		\citenamefont {Zhou}, \citenamefont {Wang}, \citenamefont {Park},
		\citenamefont {Wan},\ and\ \citenamefont {Ma}}]{Lin2021}%
	\BibitemOpen
	\bibfield  {author} {\bibinfo {author} {\bibfnamefont {G.}~\bibnamefont
			{Lin}}, \bibinfo {author} {\bibfnamefont {J.}~\bibnamefont {Jeong}}, \bibinfo
		{author} {\bibfnamefont {C.}~\bibnamefont {Kim}}, \bibinfo {author}
		{\bibfnamefont {Y.}~\bibnamefont {Wang}}, \bibinfo {author} {\bibfnamefont
			{Q.}~\bibnamefont {Huang}}, \bibinfo {author} {\bibfnamefont
			{T.}~\bibnamefont {Masuda}}, \bibinfo {author} {\bibfnamefont
			{S.}~\bibnamefont {Asai}}, \bibinfo {author} {\bibfnamefont {S.}~\bibnamefont
			{Itoh}}, \bibinfo {author} {\bibfnamefont {G.}~\bibnamefont {Günther}},
		\bibinfo {author} {\bibfnamefont {M.}~\bibnamefont {Russina}}, \bibinfo
		{author} {\bibfnamefont {Z.}~\bibnamefont {Lu}}, \bibinfo {author}
		{\bibfnamefont {J.}~\bibnamefont {Sheng}}, \bibinfo {author} {\bibfnamefont
			{L.}~\bibnamefont {Wang}}, \bibinfo {author} {\bibfnamefont {J.}~\bibnamefont
			{Wang}}, \bibinfo {author} {\bibfnamefont {G.}~\bibnamefont {Wang}}, \bibinfo
		{author} {\bibfnamefont {Q.}~\bibnamefont {Ren}}, \bibinfo {author}
		{\bibfnamefont {C.}~\bibnamefont {Xi}}, \bibinfo {author} {\bibfnamefont
			{W.}~\bibnamefont {Tong}}, \bibinfo {author} {\bibfnamefont {L.}~\bibnamefont
			{Ling}}, \bibinfo {author} {\bibfnamefont {Z.}~\bibnamefont {Liu}}, \bibinfo
		{author} {\bibfnamefont {L.}~\bibnamefont {Wu}}, \bibinfo {author}
		{\bibfnamefont {J.}~\bibnamefont {Mei}}, \bibinfo {author} {\bibfnamefont
			{Z.}~\bibnamefont {Qu}}, \bibinfo {author} {\bibfnamefont {H.}~\bibnamefont
			{Zhou}}, \bibinfo {author} {\bibfnamefont {X.}~\bibnamefont {Wang}}, \bibinfo
		{author} {\bibfnamefont {J.-G.}\ \bibnamefont {Park}}, \bibinfo {author}
		{\bibfnamefont {Y.}~\bibnamefont {Wan}},\ and\ \bibinfo {author}
		{\bibfnamefont {J.}~\bibnamefont {Ma}},\ }\bibfield  {title} {\enquote
		{\bibinfo {title} {{Field-induced quantum spin disordered state in spin-1/2
					honeycomb magnet Na$_2$Co$_2$TeO$_6$}},}\ }\href
	{https://doi.org/10.1038/s41467-021-25567-7} {\bibfield  {journal} {\bibinfo
			{journal} {Nature Communications}\ }\textbf {\bibinfo {volume} {12}},\
		\bibinfo {pages} {5559} (\bibinfo {year} {2021})}\BibitemShut {NoStop}%
	\bibitem [{\citenamefont {Kim}\ \emph {et~al.}(2021)\citenamefont {Kim},
		\citenamefont {Jeong}, \citenamefont {Lin}, \citenamefont {Park},
		\citenamefont {Masuda}, \citenamefont {Asai}, \citenamefont {Itoh},
		\citenamefont {Kim}, \citenamefont {Zhou}, \citenamefont {Ma},\ and\
		\citenamefont {Park}}]{Kim2022}%
	\BibitemOpen
	\bibfield  {author} {\bibinfo {author} {\bibfnamefont {C.}~\bibnamefont
			{Kim}}, \bibinfo {author} {\bibfnamefont {J.}~\bibnamefont {Jeong}}, \bibinfo
		{author} {\bibfnamefont {G.}~\bibnamefont {Lin}}, \bibinfo {author}
		{\bibfnamefont {P.}~\bibnamefont {Park}}, \bibinfo {author} {\bibfnamefont
			{T.}~\bibnamefont {Masuda}}, \bibinfo {author} {\bibfnamefont
			{S.}~\bibnamefont {Asai}}, \bibinfo {author} {\bibfnamefont {S.}~\bibnamefont
			{Itoh}}, \bibinfo {author} {\bibfnamefont {H.-S.}\ \bibnamefont {Kim}},
		\bibinfo {author} {\bibfnamefont {H.}~\bibnamefont {Zhou}}, \bibinfo {author}
		{\bibfnamefont {J.}~\bibnamefont {Ma}},\ and\ \bibinfo {author}
		{\bibfnamefont {J.-G.}\ \bibnamefont {Park}},\ }\bibfield  {title} {\enquote
		{\bibinfo {title} {{Antiferromagnetic Kitaev interaction in Jeff = 1/2 cobalt
					honeycomb materials Na$_3$Co$_2$SbO$_6$ and Na$_2$Co$_2$TeO$_6$}},}\ }\href
	{https://doi.org/10.1088/1361-648X/ac2644} {\bibfield  {journal} {\bibinfo
			{journal} {Journal of Physics: Condensed Matter}\ }\textbf {\bibinfo {volume}
			{34}},\ \bibinfo {pages} {045802} (\bibinfo {year} {2021})}\BibitemShut
	{NoStop}%
	\bibitem [{\citenamefont {Zhang}\ \emph {et~al.}(2023)\citenamefont {Zhang},
		\citenamefont {Xu}, \citenamefont {Halloran}, \citenamefont {Zhong},
		\citenamefont {Broholm}, \citenamefont {Cava}, \citenamefont {Drichko},\ and\
		\citenamefont {Armitage}}]{Zhang2023}%
	\BibitemOpen
	\bibfield  {author} {\bibinfo {author} {\bibfnamefont {X.}~\bibnamefont
			{Zhang}}, \bibinfo {author} {\bibfnamefont {Y.}~\bibnamefont {Xu}}, \bibinfo
		{author} {\bibfnamefont {T.}~\bibnamefont {Halloran}}, \bibinfo {author}
		{\bibfnamefont {R.}~\bibnamefont {Zhong}}, \bibinfo {author} {\bibfnamefont
			{C.}~\bibnamefont {Broholm}}, \bibinfo {author} {\bibfnamefont {R.~J.}\
			\bibnamefont {Cava}}, \bibinfo {author} {\bibfnamefont {N.}~\bibnamefont
			{Drichko}},\ and\ \bibinfo {author} {\bibfnamefont {N.~P.}\ \bibnamefont
			{Armitage}},\ }\bibfield  {title} {\enquote {\bibinfo {title} {{A magnetic
					continuum in the cobalt-based honeycomb magnet BaCo$_2$(AsO$_4$)$_2$}},}\
	}\href {https://doi.org/10.1038/s41563-022-01403-1} {\bibfield  {journal}
		{\bibinfo  {journal} {Nature Materials}\ }\textbf {\bibinfo {volume} {22}},\
		\bibinfo {pages} {58--63} (\bibinfo {year} {2023})}\BibitemShut {NoStop}%
	\bibitem [{\citenamefont {Kittel}(2004)}]{Kittel2004}%
	\BibitemOpen
	\bibfield  {author} {\bibinfo {author} {\bibfnamefont {C.}~\bibnamefont
			{Kittel}},\ }\href@noop {} {\emph {\bibinfo {title} {{Introduction to Solid
					State Physics}}}}\ (\bibinfo  {publisher} {John Wiley \& Sons, Ltd},\
	\bibinfo {year} {2004})\BibitemShut {NoStop}%
	\bibitem [{\citenamefont {Hase}, \citenamefont {Terasaki},\ and\ \citenamefont
		{Uchinokura}(1993)}]{Hase1993}%
	\BibitemOpen
	\bibfield  {author} {\bibinfo {author} {\bibfnamefont {M.}~\bibnamefont
			{Hase}}, \bibinfo {author} {\bibfnamefont {I.}~\bibnamefont {Terasaki}},\
		and\ \bibinfo {author} {\bibfnamefont {K.}~\bibnamefont {Uchinokura}},\
	}\bibfield  {title} {\enquote {\bibinfo {title} {{Observation of the
					spin-Peierls transition in linear ${\mathrm{Cu}}^{2+}$ (spin-1/2) chains in
					an inorganic compound ${\mathrm{CuGeO}}_{3}$}},}\ }\href
	{https://doi.org/10.1103/PhysRevLett.70.3651} {\bibfield  {journal} {\bibinfo
			{journal} {Phys. Rev. Lett.}\ }\textbf {\bibinfo {volume} {70}},\ \bibinfo
		{pages} {3651--3654} (\bibinfo {year} {1993})}\BibitemShut {NoStop}%
	\bibitem [{\citenamefont {Sch\"apers}\ \emph {et~al.}(2013)\citenamefont
		{Sch\"apers}, \citenamefont {Wolter}, \citenamefont {Drechsler},
		\citenamefont {Nishimoto}, \citenamefont {M\"uller}, \citenamefont
		{Abdel-Hafiez}, \citenamefont {Schottenhamel}, \citenamefont {B\"uchner},
		\citenamefont {Richter}, \citenamefont {Ouladdiaf}, \citenamefont {Uhlarz},
		\citenamefont {Beyer}, \citenamefont {Skourski}, \citenamefont {Wosnitza},
		\citenamefont {Rule}, \citenamefont {Ryll}, \citenamefont {Klemke},
		\citenamefont {Kiefer}, \citenamefont {Reehuis}, \citenamefont {Willenberg},\
		and\ \citenamefont {S\"ullow}}]{M2013}%
	\BibitemOpen
	\bibfield  {author} {\bibinfo {author} {\bibfnamefont {M.}~\bibnamefont
			{Sch\"apers}}, \bibinfo {author} {\bibfnamefont {A.~U.~B.}\ \bibnamefont
			{Wolter}}, \bibinfo {author} {\bibfnamefont {S.-L.}\ \bibnamefont
			{Drechsler}}, \bibinfo {author} {\bibfnamefont {S.}~\bibnamefont
			{Nishimoto}}, \bibinfo {author} {\bibfnamefont {K.-H.}\ \bibnamefont
			{M\"uller}}, \bibinfo {author} {\bibfnamefont {M.}~\bibnamefont
			{Abdel-Hafiez}}, \bibinfo {author} {\bibfnamefont {W.}~\bibnamefont
			{Schottenhamel}}, \bibinfo {author} {\bibfnamefont {B.}~\bibnamefont
			{B\"uchner}}, \bibinfo {author} {\bibfnamefont {J.}~\bibnamefont {Richter}},
		\bibinfo {author} {\bibfnamefont {B.}~\bibnamefont {Ouladdiaf}}, \bibinfo
		{author} {\bibfnamefont {M.}~\bibnamefont {Uhlarz}}, \bibinfo {author}
		{\bibfnamefont {R.}~\bibnamefont {Beyer}}, \bibinfo {author} {\bibfnamefont
			{Y.}~\bibnamefont {Skourski}}, \bibinfo {author} {\bibfnamefont
			{J.}~\bibnamefont {Wosnitza}}, \bibinfo {author} {\bibfnamefont {K.~C.}\
			\bibnamefont {Rule}}, \bibinfo {author} {\bibfnamefont {H.}~\bibnamefont
			{Ryll}}, \bibinfo {author} {\bibfnamefont {B.}~\bibnamefont {Klemke}},
		\bibinfo {author} {\bibfnamefont {K.}~\bibnamefont {Kiefer}}, \bibinfo
		{author} {\bibfnamefont {M.}~\bibnamefont {Reehuis}}, \bibinfo {author}
		{\bibfnamefont {B.}~\bibnamefont {Willenberg}},\ and\ \bibinfo {author}
		{\bibfnamefont {S.}~\bibnamefont {S\"ullow}},\ }\bibfield  {title} {\enquote
		{\bibinfo {title} {{Thermodynamic properties of the anisotropic frustrated
					spin-chain compound linarite PbCuSO${}_{4}$(OH)${}_{2}$}},}\ }\href
	{https://doi.org/10.1103/PhysRevB.88.184410} {\bibfield  {journal} {\bibinfo
			{journal} {Phys. Rev. B}\ }\textbf {\bibinfo {volume} {88}},\ \bibinfo
		{pages} {184410} (\bibinfo {year} {2013})}\BibitemShut {NoStop}%
	\bibitem [{\citenamefont {Okamoto}\ \emph {et~al.}(2007)\citenamefont
		{Okamoto}, \citenamefont {Nohara}, \citenamefont {Aruga-Katori},\ and\
		\citenamefont {Takagi}}]{Okamoto2007}%
	\BibitemOpen
	\bibfield  {author} {\bibinfo {author} {\bibfnamefont {Y.}~\bibnamefont
			{Okamoto}}, \bibinfo {author} {\bibfnamefont {M.}~\bibnamefont {Nohara}},
		\bibinfo {author} {\bibfnamefont {H.}~\bibnamefont {Aruga-Katori}},\ and\
		\bibinfo {author} {\bibfnamefont {H.}~\bibnamefont {Takagi}},\ }\bibfield
	{title} {\enquote {\bibinfo {title} {{Spin-Liquid State in the $S=1/2$
					Hyperkagome Antiferromagnet
					${\mathrm{Na}}_{4}{\mathrm{Ir}}_{3}{\mathrm{O}}_{8}$}},}\ }\href
	{https://doi.org/10.1103/PhysRevLett.99.137207} {\bibfield  {journal}
		{\bibinfo  {journal} {Phys. Rev. Lett.}\ }\textbf {\bibinfo {volume} {99}},\
		\bibinfo {pages} {137207} (\bibinfo {year} {2007})}\BibitemShut {NoStop}%
	\bibitem [{\citenamefont {Bera}\ \emph {et~al.}(2016)\citenamefont {Bera},
		\citenamefont {Yusuf}, \citenamefont {Kumar}, \citenamefont {Majumder},
		\citenamefont {Ghoshray},\ and\ \citenamefont {Keller}}]{Bera2016}%
	\BibitemOpen
	\bibfield  {author} {\bibinfo {author} {\bibfnamefont {A.~K.}\ \bibnamefont
			{Bera}}, \bibinfo {author} {\bibfnamefont {S.~M.}\ \bibnamefont {Yusuf}},
		\bibinfo {author} {\bibfnamefont {A.}~\bibnamefont {Kumar}}, \bibinfo
		{author} {\bibfnamefont {M.}~\bibnamefont {Majumder}}, \bibinfo {author}
		{\bibfnamefont {K.}~\bibnamefont {Ghoshray}},\ and\ \bibinfo {author}
		{\bibfnamefont {L.}~\bibnamefont {Keller}},\ }\bibfield  {title} {\enquote
		{\bibinfo {title} {{Long-range and short-range magnetic correlations, and
					microscopic origin of net magnetization in the spin-1 trimer chain compound
					${\mathrm{CaNi}}_{3}{\mathrm{P}}_{4}{\mathrm{O}}_{14}$}},}\ }\href
	{https://doi.org/10.1103/PhysRevB.93.184409} {\bibfield  {journal} {\bibinfo
			{journal} {Phys. Rev. B}\ }\textbf {\bibinfo {volume} {93}},\ \bibinfo
		{pages} {184409} (\bibinfo {year} {2016})}\BibitemShut {NoStop}%
	\bibitem [{\citenamefont {Yogi}\ \emph {et~al.}(2019)\citenamefont {Yogi},
		\citenamefont {Bera}, \citenamefont {Mohan}, \citenamefont {Kulkarni},
		\citenamefont {Yusuf}, \citenamefont {Hoser}, \citenamefont {Tsirlin},
		\citenamefont {Isobe},\ and\ \citenamefont {Thamizhavel}}]{Yogi2019}%
	\BibitemOpen
	\bibfield  {author} {\bibinfo {author} {\bibfnamefont {A.}~\bibnamefont
			{Yogi}}, \bibinfo {author} {\bibfnamefont {A.~K.}\ \bibnamefont {Bera}},
		\bibinfo {author} {\bibfnamefont {A.}~\bibnamefont {Mohan}}, \bibinfo
		{author} {\bibfnamefont {R.}~\bibnamefont {Kulkarni}}, \bibinfo {author}
		{\bibfnamefont {S.~M.}\ \bibnamefont {Yusuf}}, \bibinfo {author}
		{\bibfnamefont {A.}~\bibnamefont {Hoser}}, \bibinfo {author} {\bibfnamefont
			{A.~A.}\ \bibnamefont {Tsirlin}}, \bibinfo {author} {\bibfnamefont
			{M.}~\bibnamefont {Isobe}},\ and\ \bibinfo {author} {\bibfnamefont
			{A.}~\bibnamefont {Thamizhavel}},\ }\bibfield  {title} {\enquote {\bibinfo
			{title} {{Zigzag spin chains in the spin-5/2 antiferromagnet
					Ba$_2$Mn(PO$_4$)$_2$}},}\ }\href {https://doi.org/10.1039/C9QI00570F}
	{\bibfield  {journal} {\bibinfo  {journal} {Inorg. Chem. Front.}\ }\textbf
		{\bibinfo {volume} {6}},\ \bibinfo {pages} {2736--2746} (\bibinfo {year}
		{2019})}\BibitemShut {NoStop}%
\end{thebibliography}

%

\end{document}